\documentclass[12pt]{article}
\pdfoutput=1
\usepackage{amssymb,amsmath} 
\usepackage{graphicx}
\usepackage{epsfig} 
\usepackage{epstopdf}
\usepackage{latexsym}
\usepackage{psfrag}   
\usepackage{subfigure}  
\usepackage{booktabs}  
\usepackage{braket}
\usepackage[toc]{appendix}
\usepackage{color,soul} 
\usepackage{datetime}
\usepackage[
      colorlinks=true, 
      linkcolor=darkblue,  
      urlcolor=blue,    
      filecolor=blue,     
      citecolor=darkgreen, 
      linktocpage=true,
      pdfstartview=FitV,
      bookmarksopen=true    
      ]{hyperref}

\definecolor{darkblue}{rgb}{0.2, 0, 0.8}
\definecolor{darkgreen}{rgb}{0.2, 0.71, 0}

\numberwithin{equation}{section}

\newcommand{\req}[1]{(\ref{#1})} 
\newcommand{\labell}[1]{\label{#1}}

\newcommand{\bea}{\begin{eqnarray}}
\newcommand{\eea}{\end{eqnarray}}
\newcommand{\ba}{\begin{eqnarray}}
\newcommand{\ea}{\end{eqnarray}}
\newcommand{\nn}{\nonumber \\}

\newcommand{\beq}{\begin{equation}}
\newcommand{\eeq}{\end{equation} }
\newcommand{\beqa}{\begin{eqnarray}}
\newcommand{\eeqa}{\end{eqnarray}}
\newcommand{\beqar}{\begin{eqnarray*}}
\newcommand{\eeqar}{\end{eqnarray*}}
\newcommand{\e}{\epsilon}
\newcommand{\reef}[1]{(\ref{#1})}
\newcommand{\ssc}{\scriptscriptstyle}
\newcommand{\eg}{{\it e.g.,}\ }
\newcommand{\ie}{{\it i.e.,}\ }
\newcommand{\comment}[1]{{\bf [[#1]] }}

\newcommand{\mt}[1]{\textrm{\tiny #1}}
\newcommand{\kacs}{\kappa^{ \rm cs} }
\newcommand{\kaf}{\kappa^{ \rm f} }

\newcommand{\sigs}{\sigma^{\rm cs}}
\newcommand{\sigf}{\sigma^{\rm f}}

\newcommand{\hs}{h^{\rm cs}}
\newcommand{\hf}{h^{\rm f}}

\newcommand{\te}{t_\mt{E}}

\newcommand{\ren}{R\'enyi\ }
\newcommand{\sflat}{s_{\ssc \rm flat}}

\newcommand{\see}{S_{\rm \ssc EE}}

\newcommand{\cs}{c_{\ssc S}}

\newcommand{\ctt}{C_{\ssc T}}
\newcommand{\cthol}{C_{\ssc T}^{\rm hol}}

\newcommand{\twist}{\tau}  
\newcommand{\hyp}[1]{\mathbb H^{#1}}
\newcommand{\hF}{\widehat{\mathcal{F}}}
\newcommand{\emi}{{\rm \ssc Ext}} 
\newcommand{\reny}{R\'enyi }
\newcommand{\ato}{\tilde a^{(1)\!}}
\newcommand{\att}{\tilde a^{(2)\!}}
\newcommand{\sigtwo}{\sigma^{\rm\ssc (2)}}   
\newcommand{\rfig}[1]{Fig.\thinspace\ref{#1}}
\DeclareMathOperator{\tr}{Tr}  

\renewcommand{\href}[2]{#2}

\begin{document}  


\begin{titlepage}

 \begin{flushright}
{\tt IFT-UAM/CSIC-15-076} \\
\end{flushright}

\vspace*{2.3cm}

\begin{center}
{\LARGE \bf Universal corner entanglement}\\{\LARGE \bf  from twist operators} \\

\vspace*{1.2cm}

{\bf Pablo Bueno$^{1}$, Robert C. Myers$^{2}$ and William Witczak-Krempa$^{2}$}
\medskip

$^{1}$Instituto de F\'isica Te\'orica UAM/CSIC \\
C/ Nicol\'as Cabrera, 13-15, C.U. Cantoblanco, 28049 Madrid, Spain
\bigskip

$^{2}$Perimeter Institute for Theoretical Physics \\
31 Caroline Street North, ON N2L 2Y5, Canada 
\bigskip

\end{center}

\vspace*{0.1cm}

\begin{abstract}  
The entanglement entropy in three-dimensional conformal field theories (CFTs) receives a logarithmic contribution characterized by a regulator-independent function
$a(\theta)$ when the entangling surface contains a sharp corner with opening angle $\theta$. In the limit of a smooth surface ($\theta\rightarrow \pi$), this corner contribution vanishes as $a(\theta)=\sigma\,(\theta-\pi)^2$. In {\tt arXiv:1505.04804}, we provided evidence for the conjecture that for any 
$d=3$ CFT, this corner coefficient $\sigma$ is determined by $\ctt$, the coefficient appearing in the  two-point function of the stress tensor. Here, we argue that this
is an instance of a much more general relation connecting the analogous corner coefficient $\sigma_n$ appearing in the 
$n$th \ren entropy  
 and the scaling dimension $h_n$ of the corresponding twist operator. In particular, we find the simple relation $h_n/\sigma_n=(n-1)\pi$. 
We show how it reduces to our previous result as $n\to1$, and explicitly check its validity for free scalars
and fermions. With this new relation, we show that as $n\to 0$, $\sigma_n$ yields the coefficient of the thermal entropy, $\cs$. 
We also reveal a surprising duality relating the corner coefficients of the scalar and the fermion. 
Further, we use our result to predict $\sigma_n$ for holographic CFTs dual to four-dimensional Einstein gravity.  
Our findings generalize to other dimensions, and we emphasize the connection to the interval \ren entropies of $d=2$ CFTs.  
\end{abstract}

\end{titlepage}

\setcounter{tocdepth}{2}
{\small
\setlength\parskip{-0.5mm} 
\tableofcontents
}

\section{Introduction \& main results} 
\label{sec:Introduction} 

Understanding the structure of quantum entanglement in complex systems has become an active area of study in a variety of areas of physics, including condensed matter physics, \eg \cite{wenx,cmt}; quantum field theory, \eg \cite{cardy0,CHdir}; and quantum gravity, \eg \cite{mvr,rt0}. 
For these investigations, 
the entanglement entropy $\see$ and \ren entropies $S_n$ \cite{renyi0} have proven to be two particularly useful 
measures of the relevant degrees of freedom. In the context of quantum field theory, these are defined for a spatial region $V$ with: 
\begin{align}\label{renyi}
S_n(V)=\frac{1}{1-n}\log\, \tr \rho_V^n \,,\qquad \see(V)=\lim_{n\to1}S_n(V)=-\tr \left( \rho_V \log \rho_V \right)\,,
\end{align} 
where $\rho_V$ is the reduced density matrix computed by integrating out the degrees of freedom in the complementary region $\overline{V}$. 

\begin{figure}[h]
\center
 \subfigure[]{\label{corner}\includegraphics[scale=.7]{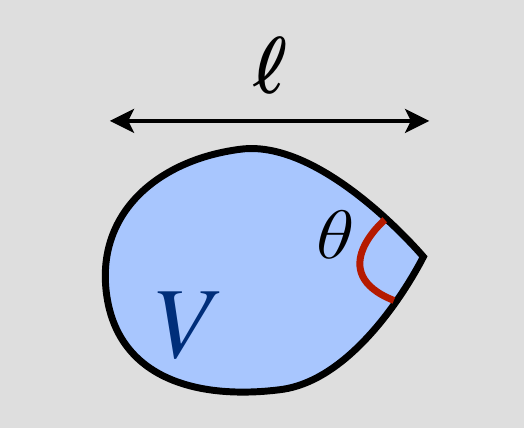}} \hspace{2cm} 
 \subfigure[]{\label{wide} \includegraphics[scale=.5]{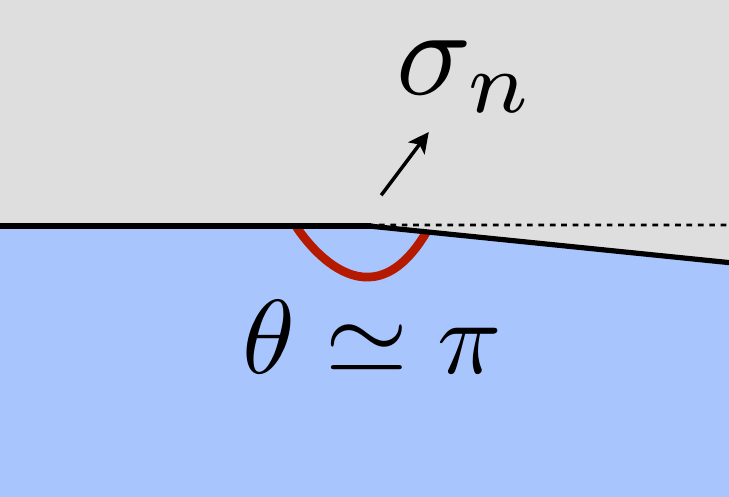}} 
\caption{a) Region $V$ whose boundary contains a sharp corner with opening angle $\theta$. b)
The contribution to the \reny entropy $S_n$ from a corner in the almost smooth limit yields a great deal of 
insight into the degrees of freedom of the CFT via the coefficient $\sigma_n$.}
\labell{corner1}
\centering
\end{figure}
In the present paper, we will focus on the \ren entropy in three-dimensional conformal field theories (CFTs), which takes the form
\begin{align}\labell{align}
S_n=B_n\frac{\ell}{\delta} - a_n(\theta) \log(\, \ell/\delta \,)+c_n+{\cal O}\big(\delta/\ell\big) \, ,
\end{align}
where $\delta$ is a UV cut-off and $\ell$ is a length scale characteristic of the size of the region $V$ --- see \rfig{corner}.  The result is dominated by the first contribution, the celebrated `area law' term, but the corresponding coefficient $B_n$ depends on the details of the UV regulator. The subleading logarithmic contribution appears when the entangling surface (\ie the boundary of $V$) contains a sharp corner of opening 
angle $\theta$, as in \rfig{corner}. The corresponding corner function $a_n(\theta)$ is regulator independent and hence it is a useful quantity to characterize the underlying CFT. 
For instance, several groups have numerically studied the corner function using lattice Hamiltonians \cite{num-corner,Laflorencie},
and obtained results independent of the lattice details which probe the low energy degrees of freedom. 

\begin{figure}
\center
 \subfigure[]{\label{an-gen}\includegraphics[scale=.55]{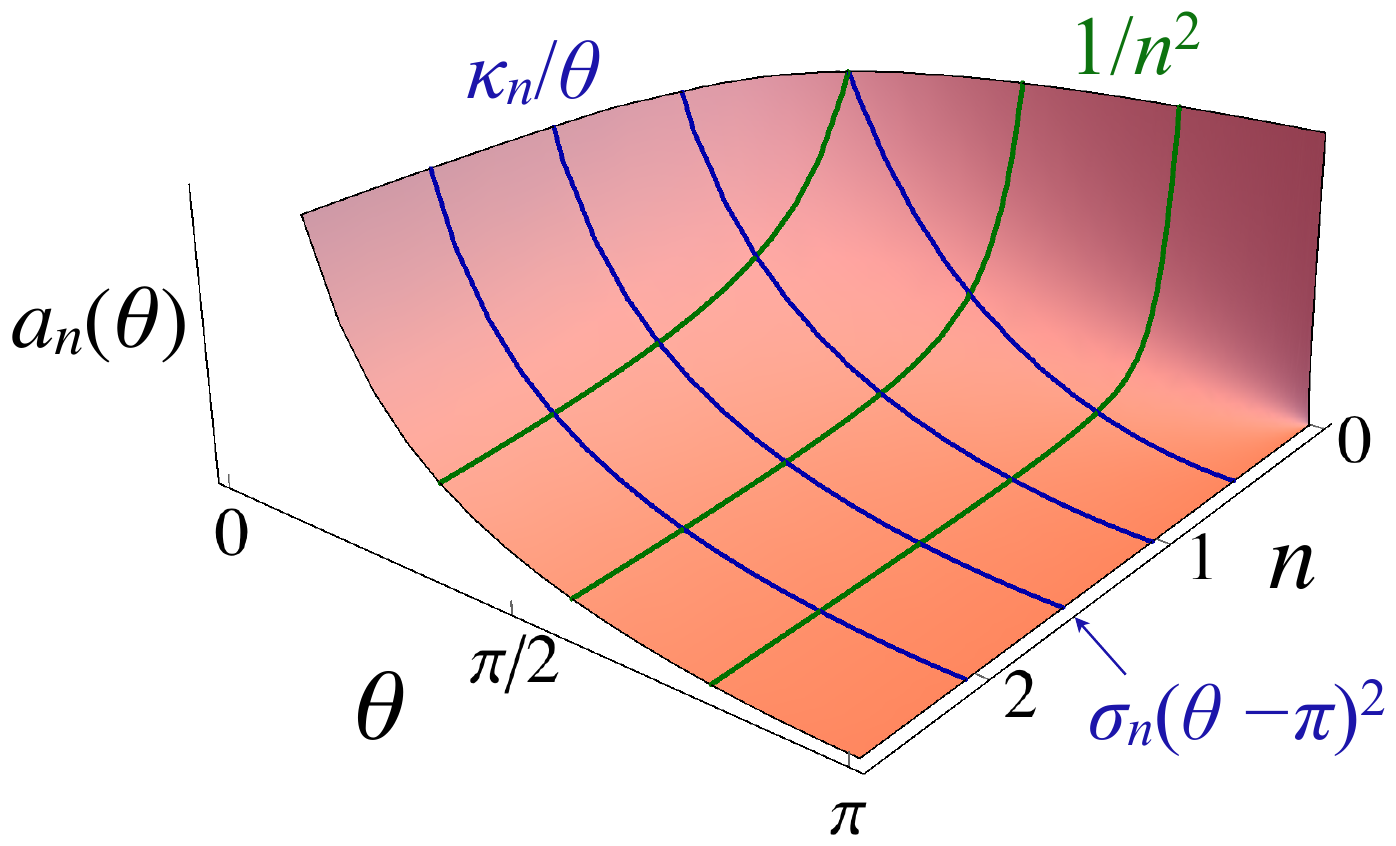}} \hspace{.8cm}  
 \subfigure[]{\label{sig-gen} \includegraphics[scale=.35]{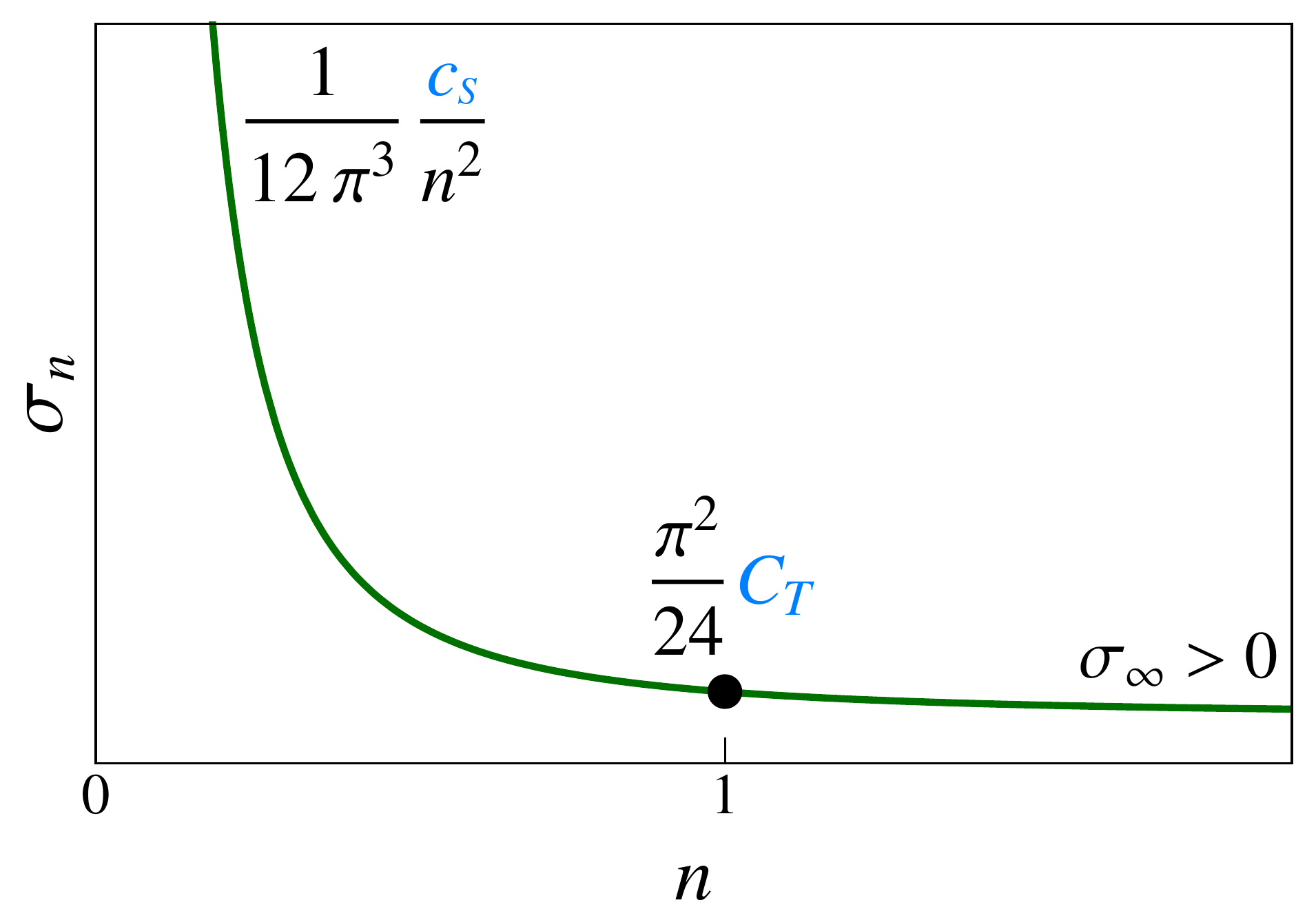}}  
\caption{Schematics for the corner entanglement function in general CFTs in $d\!=\!3$ 
spacetime dimensions. a) \ren corner contribution $a_n(\theta)$ versus the \ren index $n$ and opening angle $\theta$, 
with the asymptotics for $\theta\to0,\pi$ and $n\to0$. 
b) Smooth-corner coefficient $\sigma_n$ versus $n$. 
The divergence at small $n$ is determined by the flat-space thermal entropy density: $\sflat=\cs\, T^2$.}    
\labell{general-sig} 
\centering
\end{figure}

As the corner function $a_n(\theta)$ is central to our investigation, let us summarize a few of its key properties:
If eq.~\reef{align} is evaluated for the vacuum state, reflection positivity \cite{positive1,positive2} constrains $a_n(\theta)$ to be a positive convex function of $\theta$, \ie    
\begin{align}\labell{monotonic}
a_n(\theta)\geq 0\,,\qquad  \partial_{\theta}a_n(\theta)\leq 0\,,  \qquad \partial^2_{\theta}a_n(\theta)\geq 0\,,
\end{align}
in the range $0\le\theta\le\pi$. In fact, reflection positivity gives rise to an infinite tower 
of nonlinear higher-derivative constraints as well \cite{positive1}.\footnote{We thank Horacio Casini for explaining these points. We discuss the nonlinear constraints further in section \ref{sec:interp}.}  
Further, this function satisfies $a_n(\theta)=a_n(2\pi-\theta)$ if eq.~\reef{align} is evaluated for a pure state, \eg in the vacuum state of the CFT. 
The form of $a_n(\theta)$ is also constrained on general grounds in the limits where the corner becomes very sharp ($\theta \rightarrow 0$) and where it becomes almost smooth ($\theta\rightarrow \pi$, \rfig{wide}) \cite{CHdir}:
\begin{align}\labell{charg}
a_n(\theta\rightarrow 0)= \frac{\kappa_n}{\theta}\, ,\qquad a_n(\theta\rightarrow \pi) = \sigma_n\,(\pi-\theta)^2\, .
\end{align}
This behaviour is schematically illustrated in \rfig{an-gen}. Hence these limits define two regulator-independent coefficients, $\kappa_n$ and $\sigma_n$, which are representative of the CFT.

In studying the corner contribution in the entanglement entropy \cite{bueno1,bueno2}, we recently conjectured 
that the smooth-corner coefficient $\sigma_1$ has a universal form  in general three-dimensional CFTs, 
\begin{align}\labell{conj1}
\sigma_1&=\frac{\pi^2}{24}\,\ctt\, ,
\end{align}
where $\ctt$ is the central charge appearing in the two-point function of the stress tensor  --- see eq.~\req{t2p}.
We have verified that this relation holds for a free conformally coupled scalar and a free massless fermion, 
as well as for an eight-parameter family of strongly coupled holographic CFTs \cite{bueno1,bueno2}. A more general
holographic proof appears in \cite{rxm,prep2}.
Our primary result here is the generalization of eq.~\reef{conj1} to arbitrary values of the \ren index $n>0$:
\begin{align}\labell{conj2x}
\sigma_n = \frac{1}{\pi}\, \frac{h_n}{n-1} \, , 
\end{align}
where $h_n$ is the scaling dimension of the twist operator $\tau_n$ appearing in calculations of $S_n(V)$.
It is defined in the $n$-fold replicated theory as the surface operator on the boundary of $V$ which acts to permute 
the $n$ copies of the original QFT --- see 
section \ref{twist} for more details.   

A first test of eq.~\reef{conj2x} is to verify that we recover eq.~\reef{conj1} from this new relation in the 
limit $n\to1$. To do so, we make two observations: First, the twist operator becomes trivial at $n=1$ and hence  
the scaling dimension $h_n$ vanishes in this limit. Second, at $n=1$ the first derivative of $h_n$ with respect to $n$ is proportional to the central charge $\ctt$ in any $d$-dimensional CFT \cite{holoren,twist}. In $d=3$, the precise relation is
\begin{align}
\left.\partial_n h_n\right\vert_{n=1} &= \frac{\pi^3}{24}\, \ctt \, .
\labell{d3h}
\end{align}
Therefore in the vicinity of $n=1$, the scaling dimension can be expanded as
\beq
h_n\ \stackrel{n\to1}{=}\ \frac{\pi^3}{24}\, \ctt\, (n-1) + {\cal O}\big((n-1)^2\big)\,.
\labell{expandx}
\eeq
Now it is straightforward to see that substituting this expression into eq.~\reef{conj2x} and taking the limit $n\to1$ precisely reproduces the original relation \reef{conj1}.

Further, we have been able to verify eq.~\reef{conj2x} for a free conformally coupled scalar field and for a free massless fermion. In particular, as we discuss in the following, we can evaluate the corner coefficient $\sigma_n$ using the results of \cite{CHdir,Elvang} while the scaling dimension $h_n$ can be determined with the results of \cite{holoren,twist}. We demonstrate that these two independent calculations yield complete agreement with eq.~\reef{conj2x} for any integer values of $n>1$.  
Since the methods involved in these two calculations are so completely disparate, we find this agreement to be very strong evidence for the new conjecture. 

Before moving to detailed discussions, let us draw an interesting parallel with two-dimensional CFTs where twist operators are well understood. 
In the case of $d=2$, the twist operator is a local primary operator with scaling dimension \cite{cardy0,cardyCFT}:
\begin{align}
h^{\ssc(2)}_n=\frac{c}{12}\left(n-\frac{1}{n} \right)  \, ,
\labell{hn-twod}
\end{align}
where  $c$ is the Virasoro central charge of the theory. The \ren entropy of a single interval is calculated by first evaluating the correlator of two twist operators inserted 
at each of the endpoints and the final result can be written as
\begin{align}
  S_n^{d=2} &= 2\; \sigtwo_n \;\log(\,\ell/ \delta\,) +\dotsb \,, \qquad\quad{\rm where}\qquad
  \sigtwo_n \equiv \frac{h^{\ssc(2)}_n}{n-1}\,. 
 \labell{twod}    
\end{align} 
In this expression, $\ell$ is the length of the interval and $\delta$, the UV cut-off. The factor of two here comes from 
having two endpoints \cite{cardyCFT}.  
Hence, there is a striking similarity between the coefficient of the logarithmic contribution in $d=2$ and in $d=3$ --- in 
the limit $\theta\to\pi$ for the latter. The parallel extends to the $n\to1$ limit, where one recovers the well-known 
result $\see^{d=2}=(c/3)\,\log(\ell/\delta)$ by applying the two-dimensional analog\footnote{Eq.~\req{house} provides the general equation relating $\partial_n h_n|_{n=1}$ and $\ctt$ in $d$ dimensions. The above result follows from simply substituting $d=2$ and also $\ctt=c/(2\pi^2)$.} of eq.~\reef{d3h}: $\partial_n h_n|_{n=1}=c/6$. While it calls for some deeper physical insight, this interesting connection serves as a stepping stone from the well-known results for $d=2$ to new considerations of \ren entropies in higher dimensions --- see further discussion in sections \ref{surprize} and \ref{disc}. 

\subsection{Main results}  
Eq.~\req{conj2x} is our main result. In particular, we conjecture this to be a relation valid for general CFTs in $d=3$.
We thus predict that the corner coefficient of the R\'enyi entropies in the smooth limit $\sigma_n$ is proportional to the scaling dimension $h_n$ of the twist operator. Most of the remainder of the paper is devoted to supporting this new conjecture, 
and to extract various consequences from it.  

Eq.~\req{conj2x} incorporates our previous conjecture \req{conj1} when $n=1$, as we discussed above. Using independent computations of $\sigma_n$ \cite{CHdir,Elvang} and $h_n$  \cite{holoren,twist}, we have verified that \req{conj2x} holds exactly for a free scalar and for a free massless Dirac fermion at integer values of $n$ (up to $n=500$) as well as in the $n\rightarrow \infty$ limit. 
Further, since we have a simple expression for $h_n$ that holds for any real $n$, we can predict the behavior of $\sigma_n$ for 
non-integer $n$. This is plotted in \rfig{sighCT} for the free complex scalar and Dirac fermion. \rfig{sighCT} also includes
$\sigma_n$ for a holographic CFT dual to Einstein gravity. This is a prediction of our conjecture as $h_n$ is known for these
holographic theories, but direct access to $\sigma_n$ is currently limited to $n=1$. 
Indeed, the celebrated Ryu-Takayanagi prescription \cite{rt0} only provides a holographic description of entanglement entropy. 

A general formula for the scaling dimensions $h_n$ in dimensions $d\geq 2$ was obtained in \cite{twist} --- see \req{hn3} below. 
As we explain in section \ref{sec:free fields}, $h_n$ is expressed in terms of the thermal energy density $\mathcal{E}(T)$ of 
the CFT on the hyperbolic space $\hyp{d-1}$ (the simplest geometry with constant negative curvature). 
Using eq.~\req{conj2x}, we find a new formula for the \reny corner coefficient:
\begin{align}\labell{sig2}
\sigma_n=\frac{n}{n-1} R^{3}  \left(\mathcal{E}(T_0) -\mathcal E(T_0/n) \right) \, , 
\end{align}
where $R$ is the curvature radius of the hyperbolic plane $\hyp{2}$, which also determines the 
temperature $T_0=1/(2\pi R)$. Note that neither $\sigma_n$ nor $h_n$ depend on $R$, and that this expression
is always positive since the energy density $\mathcal E(T)$ increases with temperature.
Eq.~\req{sig2} allows for explicit calculations of the \reny corner coefficient $\sigma_n$ for any $n>0$, including non-integer values of $n$. 
It also sheds light on the physical content of $\sigma_n$. For example, following the analysis of \cite{twist}, it yields our earlier 
conjecture \reef{conj1} where $\sigma_1$ is proportional to $\ctt$, the central charge appearing in the vacuum two-point function of the stress tensor:
\begin{align} \label{t2p}
\left\langle T_{\mu\nu} (x)\, T_{\eta\kappa}(0) \right\rangle=\frac{\ctt}{x^{2d}}\,\mathcal{I}_{\mu\nu,\eta\kappa}(x)\, ,
\end{align} 
where $\mathcal{I}_{\mu\nu,\eta\kappa}(x)$ is a dimensionless tensor, whose structure is fixed by conformal symmetry \cite{petkou}.
As noted above, the desired result follows from expanding $h_n$ around $n=1$ in eq.~\reef{conj2x}.
Further, in the limit $n\to 0$, we find that the corner coefficient yields another important physical constant, the coefficient $\cs$ which controls the thermal entropy density of the CFT (in flat space), \ie $s_{\ssc\rm flat}=\cs\, T^{d-1}$. In particular, we show that eq.~\reef{sig2} yields
\begin{align}\label{cs1}
\lim_{n\to 0}\sigma_{n}= \frac{\cs}{12\pi^3} \frac{1}{n^{2}}\, .
\end{align} 
In section \ref{sec:interp} we extend this result by establishing that the corner function 
$a_n(\theta)$ diverges as $1/n^2$ for any opening angle.
In the opposite limit, \ie as $n\rightarrow \infty$, eq.~\reef{sig2} becomes
\begin{align}\label{sifi}
\sigma_{\infty}=R^3\left(\mathcal{E}(T_0)-\mathcal{E}(0) \right) > 0\, . 
\end{align}
This quantity does not seem to lend itself to an interpretation in terms of flat space observables, in contrast to the $n\to 0,1$ limits.
In section \ref{sec:interp}, we show that this asymptotic behavior extends to finite
angles, \ie $a_\infty(\theta)$ is finite for all $\theta$.
In \rfig{sig-gen}, we show the generic dependence of the $\sigma_n$ coefficient on the \ren index $n$ as deduced from eq.~\req{conj2x}. 

In Fig.~\ref{sighCT}, we have normalized the $\sigma_n$ by the corresponding central charges $\ctt$, which makes them cross at $n=1$ by eq.~\reef{conj1}. The three curves all exhibit a similar dependence on $n$. In particular, they diverge as $1/n^2$ for $n\rightarrow 0$, and asymptote to constant values for $n\rightarrow \infty$. A salient feature of \rfig{sighCT} is that the curves {\it only} coincide at $n=1$, meaning
that the ratio $\sigma_n/\ctt$ is only universal for the entanglement entropy, as was independently observed in \cite{Elvang}.  
\begin{figure}[h]
\center
   \includegraphics[scale=.48]{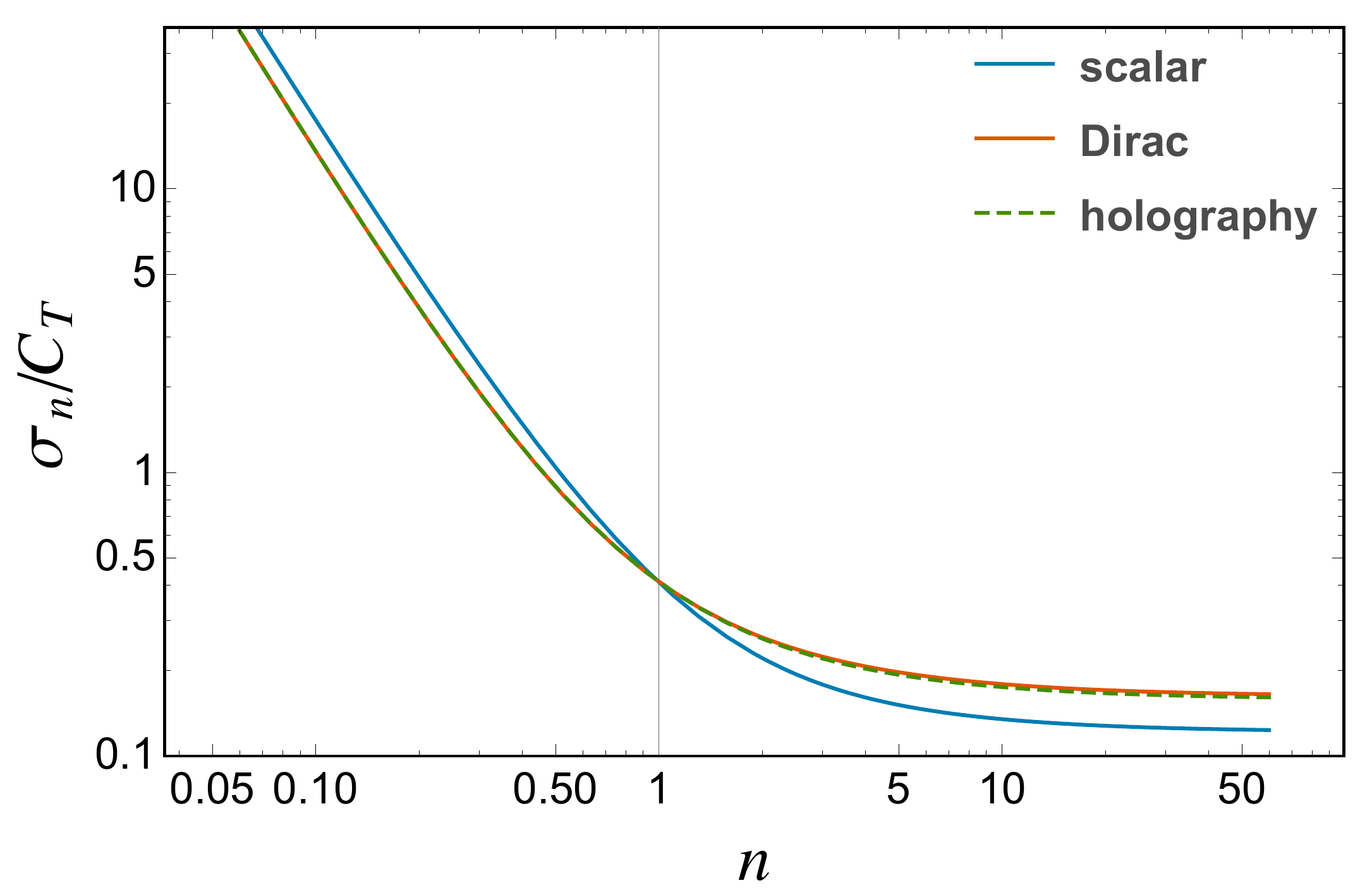}
\caption{Log-log plot of $\sigma_n/C_T$ versus the R\'enyi index $n$ for the free complex scalar, free Dirac fermion, and holographic theories.
Strikingly, the latter two are almost equal.
We note that all three lines must cross at $n=1$, where one recovers the universal ratio, $\sigma_1/C_T=\pi^2/24\simeq 0.411$.} 
\labell{sighCT}
\centering
\end{figure}  
 
Interestingly, the free fermion and holographic curves in \rfig{sighCT}  are hardly distinguishable in the whole range of values, with the agreement becoming 
even better for values $0<n\leq1$. In particular, this shows that the ratios $\cs/\ctt$ are very 
close to each other in both theories. The exact values are
\begin{align}
\frac{\cs^{\rm f}}{C_T^{\rm f}}=12\pi\zeta(3)\simeq 45.3165\, ,\qquad 
\frac{\cs^{\rm hol}}{C_T^{\rm hol}}=\frac{4\pi^5}{27}\simeq 45.3362\, .
\end{align}
While the irrational numbers involved in both cases are different, 
the final results differ by only approximately $0.04\%$.  

Our results for $\sigma_n$ also reveal a surprising duality relating the corner coefficients of the complex scalar $\sigma_n^{\rm cs}$ and fermion $\sigma_n^{\rm f}$:
\beq
n^2\,\sigma_n^{\rm cs,\,f} = \sigma_{1/n}^{\rm f,\,cs}\,, \labell{surprise0}
\eeq
which is valid for any real positive $n$.
As we show in proving this result, it requires that the corresponding thermal partition functions (on $S^1\times\hyp{2}$) are related in a relatively simple way. 
This surprising relation \reef{surprise0} also implies further connections between the two free theories, \eg the above expression can be rewritten in terms of the conformal dimensions of the corresponding twist operators, yielding $n\, h_n^{\rm cs,\,f}=-h_{1/n}^{\rm f,\,cs}$. \\ 

The remainder of the paper is organized as follows: In section \ref{twist}, we review some results concerning twist operators and their conformal dimensions $h_n$. In section \ref{sec:free fields}, we compute $h_n$ for a free conformally coupled complex scalar and a massless Dirac fermion in three dimensions. We compare these results with those obtained previously for the \reny corner coefficient $\sigma_n$, finding perfect agreement with our conjecture \reef{conj2x}. 
In this section, we also compute the coefficients $\kappa_n$, arising in the sharp corner limit ($\theta\to0$), for the free fields. These coefficients play a central
role in section \ref{sec:interp} --- see below.
Given our results for $\sigma_n$, in section \ref{surprize}, 
we discuss a surprising duality for the corner coefficients of the scalar and the fermion theories.   
In section \ref{sec:holo}, we use our conjecture to produce a new expression for $\sigma_n$ in holographic CFTs dual to Einstein gravity. 
In section \ref{sec:interp}, we study the $n$ dependence for corners of arbitrary opening angles.
Then, for a given index $n$, we propose a method for estimating the curve $a_n(\theta)$ using the coefficients $\kappa_n$ and $\sigma_n$ alone. We apply this approach to construct interpolating functions for the scalar and the fermion for $n=1,2,3$ and in the limit $n\to\infty$, and we show that they accurately fit the lattice values in the cases in which these are available. A number of technical details are provided in the appendices: 
In appendix \ref{hn_sum}, we evaluate certain integral expressions for the scaling dimensions in the free field theories to express 
$h_n$ in terms of finite closed-form sums, valid for odd $n$. Using the integral expressions for $h_n$, 
we also provide explicit values of $\sigma_n$ for certain rational values of the \ren index, which provides an interesting demonstration of the duality 
between the coefficients for the scalar and fermion theories. 
In appendix \ref{kappas}, we describe the details of calculations for the $\kappa_n$ presented in section \ref{sec:free fields}. 
Appendix \ref{revolt} explains the origin of a relation observed in \cite{Elvang} between the limit of the corner coefficient at large $n$ and the thermal partition function on $S^1\times\hyp{2}$ in the same limit. In appendix \ref{proof}, we provide a general proof of the duality between the corner coefficients of the free scalar and fermion theories. In appendix \ref{circle}, we connect the \ren corner coefficients to the \ren entropies $S_n$ for a circle, using the relation \cite{twist} between these entropies and the twist scaling dimensions $h_n$.  

\section{Twist operators} 
\label{twist}

As noted in the introduction, twist operators were originally defined in discussing \ren entropies in two-dimensional CFTs \cite{cardy0}. Twist operators are well understood in this context since they are local primary operators. They can be formally defined for quantum field theories (QFTs) in any number of dimensions, with the replica method, \eg see \cite{holoren,twist,brian1}. However, in higher dimensions, they become nonlocal surface operators and their properties are less well understood. The replica method begins by evaluating the reduced density matrix $\rho_V$ as a Euclidean path integral where independent boundary  conditions are fixed on the region $V$ as it is approached from above and below in Euclidean time, \ie with $\te\to0^\pm$. This path integral for the
partition function $Z_n$ is then extended to the Euclidean path integral on a $n$-sheeted geometry \cite{cardy0}, where the consecutive sheets are sewn together on cuts running over $V$, to represent 
\beq
\tr\! \left[\,\rho_V^{\,n}\,\right]=\frac{Z_n}{Z_1^n}\,.\labell{vbasic} 
\eeq
The denominator is introduced here to ensure the correct normalization, \ie $\tr[\rho_V]=1$.
In defining the twist operator $\twist_n[V]$, the above construction is replaced by a path integral over $n$ copies of the underlying QFT on a single copy of the geometry.  The twist operator is then defined as the codimension-two surface operator extending over the entangling surface, \ie the boundary of the region $V$, whose expectation value yields 
\begin{align}
  \langle\, \twist_n\,\rangle_n = \tr\! \left[\,\rho_V^{\,n}\,\right]\,,\labell{sloppy} 
\end{align}
where the subscript $n$ on the expectation value on the left-hand side indicates that it is taken in the $n$-fold replicated QFT. Further, here and in the following,  we omit the $V$ dependence of $\twist_n$ to alleviate the notation. Hence eq.~\reef{sloppy} implies that $\twist_n$ opens a branch cut over the region $V$ which connects consecutive copies of the QFT in the $n$-fold replicated theory. Closely related to the twist operator, 
the so-called swap operator was introduced to compute \reny entropies on the lattice \cite{swap}.  
 
In the case of a CFT, the conformal scaling dimension $h_n$ of the twist operator is defined by the coefficient of the 
leading power-law divergence in the 
correlator $\langle T_{\mu\nu}\,\twist_n\rangle_n$ as the location of $T_{\mu\nu}$ approaches that of $\twist_n$ \cite{holoren,twist}. In the case of a twist operator on an infinite (hyper)plane, as is illustrated in \rfig{twist-fig}, this correlator reads
\begin{align}
\langle T_{ab}\,\twist_n\rangle_n&=-\frac{h_n}{2\pi}\frac{\delta_{ab}}{y^d} \, , \qquad \langle T_{ai}\,\twist_n\rangle_n=0\,,\notag\\
\langle T_{ij}\,\twist_n\rangle_n&=\frac{h_n}{2\pi}\frac{(d-1)\delta_{ij}-d\,\hat n_i \hat n_j}{y^d} \, ,\labell{sing2} 
\end{align} 
where the indices $i,j$ and $a,b$ denote the two transverse directions and the $d$--2 parallel directions to the twist operator.\footnote{Let us add that implicitly the above expressions are normalized by dividing by $\langle \tau_n\rangle_n$ but we left this normalization implicit
to avoid the clutter that would otherwise be created.} Further, $y$ is the perpendicular distance from the stress tensor insertion to the twist operator and $\hat n_i$ is the unit vector orthogonally directed from $\tau_n$ to the stress tensor. 
Note that $T_{\mu\nu}$ here denotes the stress tensor for the entire $n$-fold replicated CFT. 

While the above expressions are only valid for a twist operator on a (hyper)plane, we stress that in general the leading singularity takes this form whenever $y\ll \ell$, 
where $\ell$ is any scale entering in the description of the geometry of the entangling surface. Hence the scaling dimension $h_n$ is a 
fixed coefficient  
which is characteristic of all twist operators $\tau_n$ (in a given CFT), independent of the details of the geometry of the corresponding entangling surface. Finally, let us add that $h_1=0$ since the twist operator $\twist_n$ becomes trivial for $n=1$. 

\begin{figure} 
\center
   \includegraphics[scale=.67]{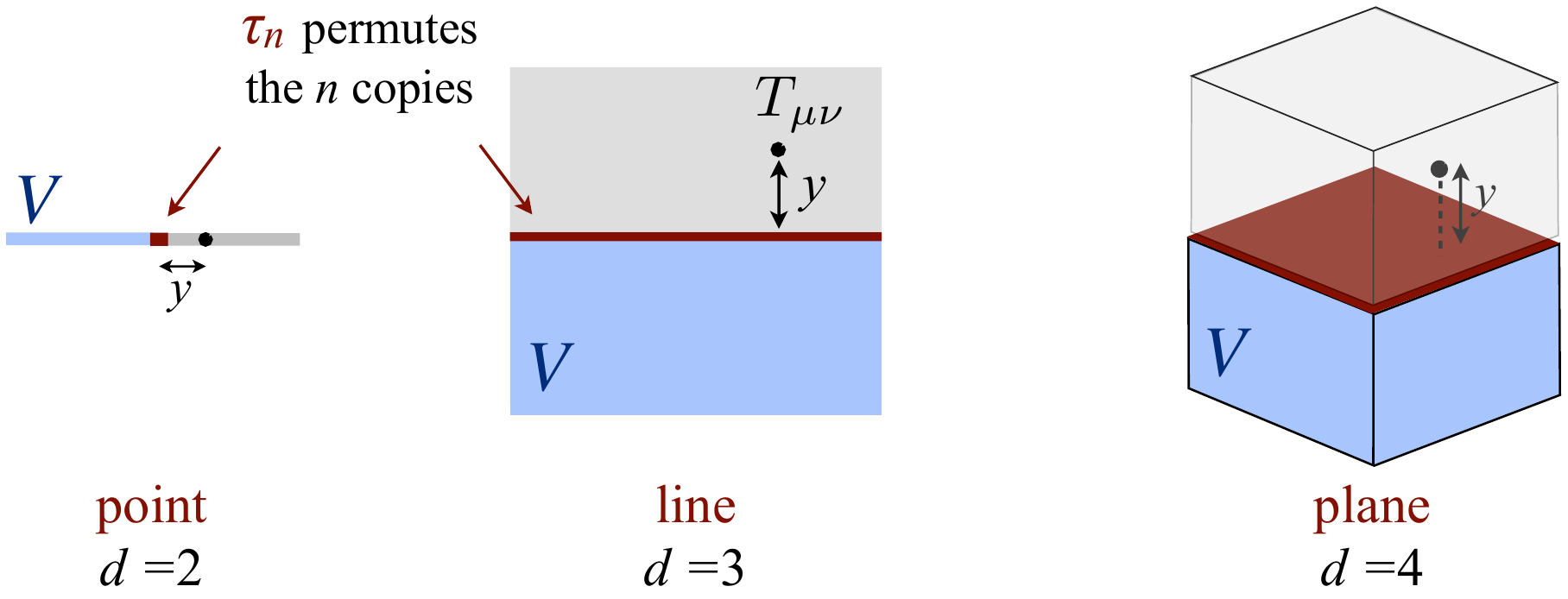}
\caption{
Planar twist operators $\tau_n$ in $d=2,3,4$ used in the definition of
the twist scaling dimension $h_n$, via eq.~\req{sing2}. They have support on the boundary (in red) of the entangling region $V$, which is
at time $t_{\rm E}=0$. The stress tensor $T_{\mu\nu}$ insertions lie a distance $y$ from the twist operators.  
In $d>2$, the use of planar boundaries here is a matter of convenience, as $h_n$ can be extracted from any entangling
surface $\partial V$ by taking $y\to 0$.
} 
\labell{twist-fig}
\centering
\end{figure}

\subsection{Insights from hyperbolic space} \labell{inside}

In refs.~\cite{holoren,CHM}, the entanglement and \ren entropies for a spherical entangling region $V=S^{d-1}$ in general CFTs were studied using a conformal mapping from the conformal vacuum in flat space to a thermal ensemble on the hyperbolic cylinder. The Euclidean version of this transformation maps a $n$-fold cover of $\mathbb R^d$ to the thermal spacetime $S^1\times \hyp{d-1}$. The flat space geometry has branch cuts beginning at a sphere of radius $R$, 
and in the hyperbolic geometry, the radius of curvature of $\hyp{d-1}$ is $R$ and the period of $S^1$ is $2\pi R\,n$. 
Using this construction, the conformal dimension $h_n$ can be expressed in terms of the thermal energy density $\mathcal{E}(T)$ 
of the (non-replicated) CFT on the hyperbolic space $\hyp{d-1}$ \cite{holoren,twist}: 
\begin{align}\labell{hn3}
h_n=\frac{2\pi\, n}{d-1}\, R^{d} \, \left(\mathcal{E}(T_0)- \mathcal{E}(T_0/n)\right)\,,
\end{align}  
where $T_0=1/(2\pi R)$. Note that the conformal dimension is independent of $R$ since this radius is the only scale in the above expression, \ie both of the energy densities above are proportional to $1/R^d$. Of course, the $R$ independence is required since $h_n$ is a dimensionless coefficient and as described above, it is independent of the geometry of the entangling surface. 

With the above definition \reef{hn3}, it is clear that setting $n=1$ yields $h_1=0$. However, as noted in the introduction, studying the limit $n\to1$ of this expression also allows one to prove \cite{twist}
\beq
\partial_nh_n|_{n=1}=2\pi^{\frac{d}{2}+1}\frac{\Gamma(d/2)}{\Gamma(d+2)}\, \ctt\,,
\labell{house}
\eeq
where $\ctt$ is the central charged appearing in the two-point correlator  \reef{t2p} of the stress tensor. 

In the $n\to 0$ limit, the temperature diverges for the second energy density 
in eq.~\reef{hn3} and this contribution dominates the result for the conformal dimension. Further, at large temperatures, the curvature of the hyperbolic space becomes unimportant and to leading order this energy density will match that in flat space \cite{brian2}. 
In particular, for a $d$-dimensional CFT, the thermal energy density in flat space directly relates to
the entropy density, $s_{\ssc \rm flat}= \cs\,T^{d-1}$, which defines a useful coefficient $\cs$ that characterizes the number of degrees of 
freedom in the CFT. The corresponding thermal energy density in flat space is then given 
by $\mathcal{E}_{\ssc\rm flat}(T)=\frac{d-1}d\, \cs\, T^d$  and hence, with 
$\mathcal{E}(T)\simeq \mathcal{E}_{\ssc\rm flat}(T)\,[1+\mathcal{O}(1/(RT)^2)]$ at high temperatures, eq.~\reef{hn3} yields
\begin{align}\labell{cs0}
\lim_{n\to0} h_{n}=-\frac{\cs}{d}\,\left(\frac{1}{2\pi\,n}\right)^{\!d-1}\, .
\end{align}
The negative sign arises here because $n<1$.
Thus, in this limit, the conformal dimension of the twist operator diverges with ($d$--1) power of the R\'enyi index, and is proportional to the flat-space thermal 
entropy coefficient. In particular, we can see this behaviour arises in two dimensions from eq.~\req{hn-twod}. The small-$n$ behaviour of the twist dimension is then
 $h_{n\to 0}^{\ssc (2)}=-c/(12n)$ and comparing with eq.~\req{cs0}, we can extract $\cs=\pi\, c/3$, which precisely agrees with the well-known result for this
 coefficient in $d=2$ \cite{cft-book}.

In the opposite limit, $n\rightarrow \infty$, the temperature vanishes in the second term in eq.~\reef{hn3}, leaving 
\begin{align}\labell{csinf}
\lim_{n\to\infty} h_{n}=\frac{2\pi\, n}{d-1}\, R^{d}\, \left(\mathcal{E}(T_0)-\mathcal{E}(0)\right)\, ,
\end{align}
which shows that $h_n$ grows linearly with $n$ when the \ren index is large, independent of the dimension $d$. 
 Let us comment here that the physical interpretation of the factor $\mathcal{E}(T_0)-\mathcal{E}(0)$ in eq.~\reef{csinf} depends on the details of the renormalization scheme used in defining the energy density $\mathcal{E}(T)$. For example, using a heat kernel regularization with free fields in the next section, we will find $\mathcal{E}(0)=0$, and hence eq.~\reef{csinf} yields $h_\infty\propto \mathcal{E}(T_0)$. However, with the holographic calculations described in section \ref{sec:holo}, one finds $\mathcal{E}(T_0)=0$ and hence in eq.~\reef{csinf}, one is left with $h_\infty\propto -\mathcal{E}(0)$. In this case, $\mathcal{E}(0)<0$ is interpreted as the Casimir energy density for the boundary CFT on the hyperbolic space. In any event, these ambiguities for the  values of the individual energy densities in eqs.~\reef{hn3} and \reef{csinf} are eliminated in the difference $\mathcal{E}(T_0)-\mathcal{E}(T_0/n)$ and so they do not affect the value of the scaling dimension of the twist operator.

\section{Free fields} 
\labell{sec:free fields}

In this section, we present explicit the free field calculations which verify our new conjecture \req{conj2x}
relating $\sigma_n$ to $h_n$. Hence in the following, we  restrict the discussion to three spacetime dimensions, in which case, $\twist_n$ is a line operator as
shown in \rfig{twist-fig}. 
We begin by computing the scaling dimensions of the twist operators $h_n$ corresponding to a free complex\footnote{Note that our discussion in \cite{bueno1} involved a real scalar, for which the values of $h_n$ and $\sigma_n$ are half those given here. We consider a complex scalar here to facilitate the comparison with the free fermion.} scalar and a free Dirac fermion in three dimensions. 
This allows us to simply compute the smooth corner coefficient $\sigma_n$ for any real $n>0$. We use this to confirm our conjecture
in the case of free CFTs, and examine some physical limits of $\sigma_n$.
In this section, we also compute the coefficients $\kappa_n$ appearing in the opposite limit, \ie that corresponding to a very sharp corner with $\theta\to0$. 
We will use the latter in section \ref{sec:interp} to estimate the value of the corner functions $a_n(\theta)$ for arbitrary values of the opening angle.

\subsection{Twist operators}  \labell{twitfree}

Recall that the scaling dimension of the twist operator $\tau_n$ in a general CFT can be evaluated with the expression in eq.~\reef{hn3}, which in three dimensions reduces to
\begin{align}\labell{hn33}
h_n=\pi\, n\, R^{3} \, \left(\mathcal{E}(T_0)- \mathcal{E}(T_0/n)\right)\,,
\end{align}  
where $\mathcal{E}(T)$ is the thermal energy density of the CFT on the hyperbolic cylinder $S^1\times\hyp{2}$. Recall that in this curved space geometry, $R$ is the radius of curvature of $\hyp{2}$, and the period of $S^1$ is set to the inverse of corresponding temperature, with $T_0=1/(2\pi R)$. Now, our first step will be to determine the partition function of the theory on the hyperbolic cylinder and then the desired thermal energy densities can be evaluated as  
\begin{align}\labell{ener}
\mathcal{E}(T)=\frac{T^2}{V_{\hyp{2}}}\,\frac{\partial }{\partial T}\log Z(T)\, ,
\end{align}
where $V_{\hyp{2}}$ is the regulated area of the hyperbolic plane $\hyp{2}$. To be precise, we have \cite{holoren,mutual}
\beq
V_{\hyp{2}}=2\pi R^2\left(\frac{R}\delta -1 + \mathcal{O}(\delta/R)\right)\,.
\labell{regulation}
\eeq
Through the conformal transformation discussed in section \ref{inside}, UV divergences in the entanglement entropy or the \ren entropy in $\mathbb R^3$ become IR divergences of the corresponding (total) thermal entropy on the hyperbolic plane $\hyp{2}$. Therefore, in order to identify quantities in a meaningful way, the UV cut-off in the flat space calculations must be mapped to the IR cut-off in the calculations on the hyperbolic cylinder. Hence, we see the short distance cut-off $\delta$ (in flat space) appearing in the regulated area above.

The partition functions for free fields on the (Euclidean) manifold $S^1\times \hyp{2}$ are readily calculated using heat kernel techniques and the results for the conformally coupled complex scalar and a massless Dirac fermion are given respectively by \cite{twist}\footnote{We emphasize that both of these expressions are UV finite (\ie the potential singularities as $x\to0$ are removed). We also note that both yield $\mathcal{E}(0)=0$ and a nonzero value for $\mathcal{E}(T_0)$ --- see comments below eq.~\reef{csinf}.} 
\begin{align}
\log Z^{\rm cs}(T)&=\frac{V_{\hyp{2}}T}{2R}\int_0^{\infty}\frac{du}{u^3\,\sinh^2 u}\,\frac{u^2+ \sinh^2 u\,(u \,\coth u-2)}{\sinh (u/(2\pi R T))}\, ,
\labell{part2}\\ \notag
\log Z^{\rm f}(T)&=\frac{V_{\hyp{2}}T}{2R}\int_0^{\infty}\frac{du}{u^3\,\sinh u}\frac{2\sinh u -u \left(u  \coth u+1 \right)}{\tanh (u/(2\pi R T))}\, ,
\end{align}  
where again $R$ is the radius of curvature of the hyperbolic plane and the temperature $T$ corresponds to the inverse period of the circle $S^1$.
It is now straightforward to evaluate the scaling dimension \req{hn3} of the corresponding twist operators using eq.~\req{ener} and  the final expressions can be written as
\begin{align}
h_n^{\rm cs}&=\frac{1}{4\pi n^2}\int_0^{\infty}\frac{du}{\sinh u}\left[\frac{\cosh(u/n)}{\sinh^3(u/n)}-n^3\,\frac{\cosh u}{\sinh^3u} \right]\, ,
 \labell{confw}\\ \notag
h_n^{\rm f}&=\frac{1}{8\pi n^2}\int_0^{\infty}\frac{du}{\tanh u}\left[n^3\,\frac{2+\sinh^2 u}{\sinh^3u}
-\frac{2+\sinh^2 (u/n)}{\sinh^3(u/n)} \right]\, . 
\end{align}
In appendix \ref{hn_sum}, we show that these integrals can be evaluated for odd values of $n$, and expressed in terms of sums involving trigonometric functions. In any event, the above expressions can be easily evaluated for different values of $n$, and we will use this to test the conjecture 
\req{conj2x} relating $h_n$ to the corner coefficient $\sigma_n$
in the next subsection. Some explicit results are collected in Table \ref{tbl1}.
\begin{table*} 
  \centering
  \begin{tabular}{c||c|c|c|c|c|c|c|c} 
  $n $  &$n\to0$ & 1+$\epsilon$&   2 & 3 & 4 & 5 & 6& $n\to\infty$\\
    \hline\hline \rule{0pt}{1.5em}
  $ \hs_n$  &$-\frac{\zeta(3)}{4\pi^3}\frac{1}{n^2}$ &$\frac{\pi}{128}\epsilon$ &   $\frac{1}{24\pi}$ & $\frac{1}{27\sqrt{3}}$ & $\frac{3\pi+8}{192\pi}$ & $\frac{\sqrt{25-2\sqrt{5}}}{125} $& $\frac{81+34\sqrt{3}\pi}{1944\pi}$& $\frac{3\zeta(3)}{16\pi^3}n$\\ \hline \rule{0pt}{1.5em}
   $\hf_n$  &$-\frac{3\zeta(3)}{16\pi^3}\frac{1}{n^2}$&$\frac{\pi}{128}\epsilon$  &   $\frac{1}{64}$ & $\frac{5}{108\sqrt{3}}$ & $\frac{1+6\sqrt{2}}{256}$ & $\frac{\sqrt{425+58\sqrt{5}}}{500} $ & $\frac{261+20\sqrt{3}}{5184}$ & $\frac{\zeta(3)}{4\pi^3}n$
  \end{tabular}
  \caption{Scaling dimensions $h_n$ of the twist operator at various values of $n$
for a massless complex scalar and a massless Dirac fermion. In the third column $|\e|\ll 1$. For non-integer $n$, see Table \ref{tbl-rational}.}
\label{tbl1}   
\end{table*}  

Following our general discussion above, let us comment further on some of the results in the table. 
First, we consider the limit $n\rightarrow 0$ for which eq.~\reef{cs0} becomes
\begin{align}\labell{cs0x}
\lim_{n\to0} h_{n}=-\frac{\cs}{12\pi^2}\,\frac{1}{n^2}\, .
\end{align}
By evaluating the integrals for small $n$, we find that our free field results have the desired form, with both of the scaling dimensions diverging as $1/n^2$ --- see Table \ref{tbl1}. Further, examining the overall coefficient of these divergences, eq.~\req{cs0x} indicates that
\begin{align}
\cs^{\rm cs}=\frac{3\zeta(3)}{\pi}\qquad{\rm and} \qquad \cs^{\rm f}=\frac{9\zeta(3)}{4\pi}\,,
\end{align}
for the thermal entropy coefficient of the complex scalar and the Dirac fermion, respectively. One can easily verify that these values precisely match the expected thermal entropy coefficients for both fields, \eg see \cite{ope}. Another interesting limit to consider is $n\rightarrow 1$. While the scaling dimension vanishes at precisely $n=1$, evaluating eq.~\req{confw} in the vicinity of this point, we find
\begin{align}\labell{lim1}
\left.h^{\rm cs}_n\right|_{n=1+\epsilon}&= \frac{\pi}{128}\,\epsilon-\frac{17 \pi }{1920}\,\epsilon^2+\mathcal{O}(\epsilon^3)\, ,\\ \notag
\left.h^{\rm f}_n\right|_{n=1+\epsilon}&=  \frac{\pi}{128}\,\epsilon-\frac{13 \pi }{1920}\,\epsilon^2+\mathcal{O}(\epsilon^3)\, .
\end{align}
Now recall that eq.~\reef{house} determined the leading coefficient in terms of the central charge $\ctt$. In particular, we have $\partial_n h_n\vert_{n=1}=\frac{\pi^3}{24}\ctt$ for $d=3$. Now given $\ctt^{\rm cs,\, f}=3/(16\pi^2)$ \cite{petkou}, this latter result yields $\partial_n h_n\vert_{n=1}=\pi/128$ for both fields, in agreement with the above expansions. Finally, the scaling dimension exhibits the expected linear growth with the \ren index as $n\rightarrow \infty$. In particular, we find:
\begin{align}\labell{hinf}
\lim_{n\rightarrow \infty} h^{\rm cs}_{n} =\frac{3\zeta(3)}{16\pi^3}\, n\qquad{\rm and} \qquad
\lim_{n\rightarrow \infty} h^{\rm f}_{n} = \frac{\zeta(3)}{4\pi^3}\,n \, .
\end{align}

\subsection{Corner R\'enyi entropies}  \labell{gamble} 
The regulator independent contributions to the entanglement and \ren entropies produced by a sharp corner in the entangling surface for $d=3$ free conformally coupled scalars and free Dirac fermions were studied in a series of papers by Casini and Huerta \cite{CHdir}. However, the expressions for, \eg the \ren coefficients $\sigma_n$ were left in a very complicated and implicit form. Motivated by our original conjecture \req{conj1} 
relating $\sigma_1$ and $\ctt$, and the possibility that a similar universal relation could also exist for $\sigma_n$, 
the results for $\sigma_n^{\rm cs,\, f}$ were recently reduced to simple and closed-form expressions 
at integer $n$ \cite{Elvang}. 
These new expressions were used to easily evaluate the first coefficients with $n\geq 1$, as well as determine the $n\rightarrow \infty$ 
behaviour. In this section, we use these results to confirm our new conjecture, verifying that eq.~\req{conj2x} holds for both the 
scalar and the fermion.  

In the second part of this section, we consider the coefficients $\kappa^{\rm cs,\, f}_n$, which control the corner contribution of the \ren entropy in the limit $\theta\to0$. In particular, we evaluate the corresponding expressions for $n=2,3,4$, and at large $n$ --- see also appendix \ref{kappas}. 
In section \ref{sec:interp}, we will use these coefficients, along with the $\sigma_n^{\rm cs,\, f}$, to construct simple interpolating functions to approximate the corner function $a^{\rm cs,\, f}_n(\theta)$ for all angles.

\subsubsection{Smooth surface limit}

The expressions for $\sigma_n^{\rm cs,\, f}$ appearing in \cite{CHdir} were recently shown to reduce to the following simple sums (for integer $n$) \cite{Elvang}: 
\begin{eqnarray}
\sigs_n=\sum_{k=1}^{n-1}\frac{k(n-k)(n-2k)\tan\left(\frac{\pi k}{n} \right)}{12\pi n^3(n-1)}  \, ,  \qquad
\sigf_n= \sum_{k=-(n-1)/2}^{(n-1)/2}\frac{k(n^2-4k^2)\tan\left(\frac{\pi k}{n} \right)}{24\pi n^3(n-1)} \,  .
\labell{sumsig}
\end{eqnarray}
Now combining eq.~\req{hn33} for the conformal dimension $h_n$ with our new conjecture \req{conj2x}, we find a new general formula for the corner coefficient in terms of the thermal energy density on the hyperbolic cylinder:
\begin{align}\labell{sin33}
\sigma_n= \frac{n}{n-1}\, R^{3} \, \left(\mathcal{E}(T_0)- \mathcal{E}(T_0/n)\right)\,.
\end{align}  
In particular, focusing on the free fields, we can use the expressions \req{confw} for $h_n$ to produce the following prediction for the corner coefficients in these theories
\begin{align}
\sigma_n^{\rm cs}&=\frac{1}{4\pi^2 n^2(n-1)}\int_0^{\infty}\frac{du}{\sinh u}\left[\frac{\cosh(u/n)}{\sinh^3(u/n)}-n^3\,\frac{\cosh u}{\sinh^3u} \right]\, ,
 \labell{confs}\\ \notag
\sigma_n^{\rm f}&=\frac{1}{8\pi^2 n^2(n-1)}\int_0^{\infty}\frac{du}{\tanh u}\left[n^3\,\frac{2+\sinh^2 u}{\sinh^3u}
-\frac{2+\sinh^2 (u/n)}{\sinh^3(u/n)} \right]\, . 
\end{align}
We have  explicitly evaluated these integrals for a few values of $n$ in Table \ref{tbl1}, and also
numerically plotted the coefficients for continuous $n$ in \rfig{sighCT}.
In appendix \ref{hn_sum}, we also evaluate the integrals analytically for odd $n>1$ to express these corner coefficients as certain closed sums --- see eqs.~\reef{sigw4}
and \reef{sigw4a}. Our new closed-form sums agree with eq.~\reef{sumsig} for all odd integer values that 
we have checked, $1<n\leq 500$, but take a very different form from the latter. It would be desirable to explicitly prove
that both are equal for all odd integers.
In any event, we have also verified that the above integrals \req{confs} exactly reproduce the values obtained from eq.~$\req{sumsig}$ for $n=1,2,\cdots,500$, as well 
as in the limit $n\to \infty$. In that limit, we have also verified that the subleading terms agree, as we discuss below.
We find that this agreement provides strong evidence for our conjecture \req{conj2x} relating the corner coefficient $\sigma_n$ and the scaling dimension $h_n$ of the corresponding twist operators.   
\begin{table}
  \centering
  \begin{tabular}{c||c|c|c|c|c|c|c|c} 
  $n $  & $n\to0$&  1 & 2 & 3 & 4 & 5 & 6& $n\to\infty$\\
    \hline\hline \rule{0pt}{1.5em}
  $ \sigs_n$  & $\frac{\zeta(3)}{4\pi^4}\, \frac{1}{n^2}$ &  $\frac{1}{128}$ & $\frac{1}{24\pi^2}$ & $\frac{1}{54\sqrt{3}\pi}$ & $\frac{3\pi+8}{576\pi^2}$ & $\frac{\sqrt{25-2\sqrt{5}}}{500\pi} $& $\frac{81+34\sqrt{3}\pi}{9720\pi^2}$& $\frac{3\zeta(3)}{16\pi^4}$ \\ \hline \rule{0pt}{1.5em}

   $\sigf_n$   & $\frac{3\zeta(3)}{16\pi^4}\, \frac{1}{n^2}$&  $\frac{1}{128}$ & $\frac{1}{64\pi}$ & $\frac{5}{216\sqrt{3}\pi}$ & $\frac{1+6\sqrt{2}}{768\pi}$ & $\frac{\sqrt{425+58\sqrt{5}}}{2000\pi} $ & $\frac{261+20\sqrt{3}}{25920\pi}$ & $\frac{\zeta(3)}{4\pi^4}$
  \end{tabular} 
  \caption{Corner coefficient $\sigma_n$ for various \reny indices for the complex scalar and Dirac CFTs in $d=3$. 
}
\labell{tbl2}   
\end{table}  

Of course, we can also use this agreement to combine the relation $h_n=\pi(n-1)\sigma_n$ with eq.~\req{sumsig} to express the scaling dimensions $h^{\rm cs\,,f}_n$ in terms of sums:
\begin{eqnarray}
h^{\rm cs}_n=\sum_{k=1}^{n-1}\frac{k(n-k)(n-2k)\tan\left(\frac{\pi k}{n} \right)}{12 n^3}  \, ,  \qquad
h^{\rm f}_n= \sum_{k=-(n-1)/2}^{(n-1)/2}\frac{k(n^2-4k^2)\tan\left(\frac{\pi k}{n} \right)}{24 n^3} \,  ,
\labell{sumh}
\end{eqnarray}
which is valid for all integer values of $n>1$.

One advantage of the integral expressions \req{confs} for $\sigma_n^{\rm cs\,, f}$ over the sums in eq.~\req{sumsig} is that they 
allow us to evaluate the corner coefficients for non-integer values of $n$. 
For example, we can examine how these coefficients vary in the vicinity of $n= 1$, \ie
\begin{align}\labell{lim11}
  \sigma^{\rm cs}_n\big|_{n=1+\epsilon}&= \frac{1}{128}-\frac{17 }{1920}\,\epsilon+\mathcal{O}(\epsilon^2)\, ,\\ \notag 
  \sigma^{\rm f}_n\big|_{n=1+\epsilon}&=  \frac{1}{128}-\frac{13  }{1920}\,\epsilon+\mathcal{O}(\epsilon^2)\, .
\end{align}
Hence we see that the agreement between the two coefficients at $n=1$ does not extend beyond this precise point, as expected from eq.~\reef{lim1}. We can also consider the limit $n\rightarrow 0$, which yields
\begin{align}\label{s0}
\lim_{n\to0} \sigma_{n}^{\rm cs}\simeq \frac{\zeta(3)}{4\pi^4}\, \frac{1}{n^2}\,, \qquad {\rm and}\qquad
\lim_{n\to0} \sigma_{n}^{\rm f}\simeq\frac{3\zeta(3)}{16\pi^4}\, \frac{1}{n^2}\,.
\end{align}
Of course, these results are simply related to the corresponding limit of the scaling dimensions in eq.~\reef{cs0x}, and hence the corner coefficients are proportional to $\cs$ and diverge as $1/n^2$ in this limit. 

Next, we note that while the $h_n$ diverge 
linearly with $n$ as $n\to\infty$ in eq.~\reef{hinf}, the $\sigma_n$ asymptote to the constant values $\sigma_\infty^{\rm cs,\, f}$ 
given in Table \ref{tbl2}. The detailed large-$n$ asymptotic behavior of both corner coefficients can be determined from eq.~\req{confs} and 
reads:\footnote{\label{fo5} We note that for the scalar, the large-$n$ expansion can be performed directly on the corresponding integral in eq.~\req{confs}. However, the analogous calculation for the fermion does not produce a consistent expansion. Rather in this case, we took advantage of an alternative representation of $\sigma_n^{\rm f}$ which makes use of the expression for $\log Z^{\rm f}(T)$ given in \cite{igor5} --- see eq.~\reef{woow} and appendix \ref{proof}.}    
\begin{align}
\lim_{n\to\infty}\sigma_n = \sigma_\infty\, \left(1+\frac{1}{n} \right)\, \left[ 1+\frac{b_2}{n^2} - \frac{b_4}{n^4}+\mathcal O\!\left( \frac{1}{n^6} \right) \right] \,,
\labell{free-large-n}
\end{align}   
where 
\beqa
b_2^{\rm cs}&=&1\,, \qquad
b_4^{\rm cs}=-1 + \frac{\pi^4}{45 \zeta(3)}\simeq 0.8008\,,
\nonumber\\
 b_2^{\rm f}&=&1 - \frac{\pi^2}{12 \zeta(3)}\simeq 0.3158\,, 
 \labell{largenX}\\
b_4^{\rm f}&=&-1 + \frac{\pi^2 (120 + 7 \pi^2)}{1440 \zeta(3)}\simeq 0.0781\,.
\nonumber
\eeqa
We further find that the series in brackets only
contains even powers of $1/n$, with terms that alternate in sign. 
These analytical results for the asymptotics agree with the numerical analysis presented in \cite{Elvang} of the sums in eq.~\req{sumsig}. This analysis yielded precise estimates for $b_2^{\rm cs,\, f}$, which coincide with the values given above. Based on their numerical fits, the authors of \cite{Elvang} conjectured
that the terms in the large-$n$ expansions appear in pairs with identical coefficients, which leads to the overall factor of $1+1/n$ in eq.~\reef{free-large-n}. Here the large-$n$ expansion of our integral expressions allow us to confirm the appearance of this factor analytically. As an aside, we observe that the same factor $1+1/n$ fully determines
the $n$ dependence of the universal coefficient of the $d=2$ \ren entropy in eq.~\req{twod}.
Overall, the large-$n$ asymptotic analysis discussed above provides further support for our main conjecture \req{conj2x}
relating $h_n$ and $\sigma_n$. 

Ref.~\cite{Elvang} further observed that for both of the free fields
\begin{align}\labell{helv}
\lim_{n\to\infty} \sigma_{n}=\frac{1}{4\pi^2} \,\mathcal{F}_\infty \,,
\end{align}
where the coefficient $\mathcal{F}_\infty$ is related to the logarithm of the partition function on $S^1\times \hyp{2}$, as calculated in \cite{igor5}. Given the connection
\reef{conj2x} between the corner coefficient and the scaling dimension, the origin of the above relation can be traced to the facts that the scaling dimension can be expressed in terms of the energy density on the hyperbolic geometry $\hyp{2}$, as in eq.~\reef{hn33}, and that the energy density is determined by a derivative of $\log Z(T)$ with respect to the temperature, as in eq.~\reef{ener}. We defer the precise details to appendix \ref{revolt} because producing the exact result in eq.~\reef{helv} requires taking into account differences between our present conventions and those in \cite{igor5}. Our analysis in the appendix establishes that eq.~\reef{helv} is a general relation that holds for any three-dimensional CFT.

\subsubsection{Sharp corner limit}   \labell{scl}

The main goal of our paper is to establish the relation \req{conj2x} between the scaling dimensions of twist operators and 
the corner coefficients $\sigma_n$, arising for almost smooth corners. However, it is also  
interesting to study the opposite limit of the corner function $a_n(\theta)$, corresponding to a very small opening angle, 
\ie $\theta\to0$. In this limit, the corner contributions are characterized by the coefficients $\kappa_n$ defined by 
\begin{align}
a_n(\theta\rightarrow 0)=\kappa_n/\theta\, .
\end{align}
We will evaluate the first few $\kappa_n^{\rm cs,\, f}$ here and use the results to produce simple functions to approximate $a_n^{\rm cs,\, f}(\theta)$ for all angles, in section \ref{sec:interp}. Further, we will explicitly show that these simple functions provide an excellent approximation to the exact $a_n(\theta)$ for $n=1,2,3$.  

Interestingly, in the case of free fields, the $\kappa_n$ coefficients are related to the so-called ``entropic c-functions'' for a family of corresponding massive free fields in $d=2$ \cite{CHdir}. In particular, they can be evaluated as
\begin{align}
\kappa_n^{\rm cs,\,  f}=\frac{1}{\pi}\int_0^{\infty}\!dt\, c^{\rm cs,\,  f}_n(t)\, ,
\end{align}
where the $c^{\rm cs,\,  f}_n(t)$ are the universal functions
\begin{align}
c^{\rm cs,\,  f}_n(m\ell)=\ell\,\frac{dS_n(m\ell)}{d\ell}\,,
\end{align}
where $S_n(m\ell)$ are the \reny entropies of $d=2$ free fields with mass $m$ for a single interval of length $\ell$. We leave the details of the calculations of the $c^{\rm cs,\,  f}_n(t)$ to appendix \ref{kappas}. The final expressions for the coefficients $\kappa_n^{\rm cs,\,  f}$ can be written as
\begin{align}\label{kas}  
\kacs_{n}&=\frac{1}{\pi(n-1)}\sum_{k=1}^{n-1}\int_0^{\infty}\!dt\,  \frac{t^2{u^{\prime}}_{k/n,\,  \rm cs}^2-{u}_{k/n,\,  \rm cs}^2((1-2k/n)^2+t^2(1+{u}_{k/n,\,  \rm cs}^2))}{2(1+{u}_{k/n,\,  \rm cs}^2)} \, ,\\ \label{kaf}
\kaf_{n}&=\frac{2}{\pi(n-1)}\sum_{k>0}^{(n-1)/2}\int_0^{\infty}\!dt\,  \frac{t^2{u^{\prime}}_{k/n,\,  \rm f}^2-{u}_{k/n,\,  \rm f}^2(4(k/n)^2+t^2(1-{u}_{k/n,\,  \rm f}^2))}{2(1-{u}_{k/n,\, \rm f}^2)}\, ,
\end{align}
which applies for integer $n>1$. The functions $u_{k/n,\rm cs,\, f}(t)$ are solutions to the differential equations \req{eqs} and \req{eqf} given in the appendix, subject to the boundary conditions \req{bds} and \req{bdf} in each case. Evaluating the expressions in eqs.~\req{kas} and \req{kaf} for $n=2,3,4,\infty$, we obtain the numerical results shown in Table \ref{tbl3}.

As also described in the appendix, $\kappa^{\rm cs,\,f}$ corresponding to the special case $n=1$ are evaluated using a similar approach using two-dimensional entropic c-functions. The values of these coefficients were evaluated by \cite{CHdir} and also appear in Table~\ref{tbl3} under $n=1$. 

\begin{table}
  \centering
  \begin{tabular}{c||c|c|c|c|c} 
  $n $  & 1 &  2 & 3 & 4 & $n\to\infty$ \\
    \hline\hline \rule{0pt}{1.5em}
  $ \kacs_n$ & 0.0794 & 0.0455996(1) & 0.037339(1)& 0.033798(1) & 0.0262(2) \\ \hline \rule{0pt}{1.5em}

   $\kaf_{n}$ & 0.0722 & 0.0472338(1)& 0.040662(1) & 0.0376674(1) & 0.030(4)
  \end{tabular} 
  \caption{Sharp-corner coefficient for various \reny indices for the massless complex scalar and massless Dirac fermion in $d=3$. }
\label{tbl3}   
\end{table}

\subsection{$n$ and  $\theta$ factorization?}    
 
One can ask whether the \ren index $n$ and $\theta$ dependences of the corner function $a_n(\theta)$ factorize, namely,
\begin{align} \label{factor}
  a_n(\theta) \stackrel{?}{=} f(n)\, a_1(\theta)\,,
\end{align}
where $f(n\!=\!1)=1$ and $a_1(\theta)$ is the corner function appearing at $n=1$
(\ie in the entanglement entropy). Such factorization occurs for the corner contribution in the extensive mutual information model \cite{brian1,CH4,CH5}, which can be written as
\begin{align} \labell{emi}
  a_n^{\emi}(\theta) = \frac{1}{\partial_nh_n|_{n=1}}\,\frac{h_n}{n-1} \,a_1^{\emi}(\theta)\,, \qquad{\rm where}\ \ 
  a_1^{\emi}(\theta)=\frac{\pi^2}8\,\ctt\left[1+(\pi-\theta)\cot\theta \right]\,.
\end{align}
This result follows from a straightforward generalization of our calculations for $n\!=\!1$ in \cite{bueno1}. We note that the precise
form of $h_n$ is undetermined for this model,
apart from the general constraints valid for any CFT.
Factorization also occurs in a more concrete setting in a class of quantum critical Lifshitz theories with dynamical exponent $z=2$,
which have a corner function characterizing their corner entanglement just like CFTs. 
Moore and Fradkin \cite{shitz} indeed found that $f(n)\equiv 1$,
\ie the corner function $a_n(\theta)$ is independent of $n$ for all angles\footnote{We thank S.\ Furukawa for pointing this out.}$^,\,$\footnote{It has been numerically shown that non-analycities can arise in the $n$-dependence 
of the corner function of certain Lifshitz QCPs when these are studied using lattice realizations \cite{joel,stephan}. However, there is no evidence that such singular effects arise in the ideal (fixed point) field theories 
describing those QCPs \cite{shitz}. 
We thank Joel Moore for bringing this point to our attention.}.  

Using our above results, we find that the factorization \req{factor} does not hold for the free complex scalar or for the Dirac fermion.
Indeed, if eq.~\req{factor} were to hold, we would have the equality $\kappa_n/\kappa_1 = \sigma_n/\sigma_1$. 
It is sufficient to examine the $n=2$ case to find a contradiction:
\begin{align} \labell{goog}
  \kacs_2/\kacs_1 &= 0.574\,, \qquad \sigs_2/\sigs_1 = 0.540\,, \nn
  \kaf_2/\kaf_1 &= 0.654\,, \qquad \ \, \sigf_2/\sigf_1 = 0.637\,. 
\end{align}
The deviations from factorization here are relatively small, \ie about 6\% for the complex scalar and 3\% for the fermion. However, 
these deviations are much greater than the numerical uncertainties in our evaluation of the corresponding $\kappa_n$ coefficients
--- recall that the $\sigma_n$ are exact. 

The failure of the factorization here is reminiscent of a similar non-factorization found in \cite{holoren} for the \ren entropies of spherical regions in higher dimensional CFTs. There the authors considered the question of whether or not the coefficient of the universal contribution in these \ren entropies took the form $f_d(n)\, a_d^*$, where $a_d^*$ is a constant determined by the specific data of the underlying CFT but $f_d(n)$ is some function of the \ren index which is the same for any CFT in $d$ dimensions. This kind of factorization is found in the \ren entropy of a single interval in two dimensions, as can be inferred from eqs.~\reef{hn-twod} and \reef{twod}. However, by examining a variety of holographic models, ref.~\cite{holoren} found that this factorization fails in any number of spacetime dimensions beyond two.

\section{Bose-Fermi duality}   \labell{surprize}  
In this section, we study duality relations between the universal entanglement coefficients at \ren indices $n$ and $1/n$, in both $d=2$ and $3$, the latter being our main focus.
First, let us make some observations comparing the corner coefficients $\sigma_n$ of the complex scalar and the Dirac fermion in $d=3$.  
We note that the ratio $\sigma^{\rm cs}_{n}/\sigma^{\rm f}_{n}$ monotonically decreases as $n$ runs from $0$ to $\infty$, \ie 
\begin{align}\labell{sinf}
\frac{\sigma^{\rm cs}_{n\rightarrow 0}}{\sigma^{\rm f}_{n\rightarrow 0}} =\frac{4}{3}\, , \qquad \frac{\sigma^{\rm cs}_{1}}{\sigma^{\rm f}_{1}} = 1\, , \qquad \frac{\sigma^{\rm cs}_{\infty}}{\sigma^{\rm f}_{\infty}} =\frac{3}{4}\, .
\end{align}
Interestingly, the values corresponding to the complex scalar and the Dirac fermion are closely related in the opposite limits
of small and large $n$:
\begin{align}\labell{sinf}
  \left.n^2\,\sigma_n^{\rm cs} \right\vert_{n\rightarrow 0}=\sigma^{\rm f}_{\infty}\, , \qquad 
  \left.n^2\,\sigma_n^{\rm f} \right\vert_{n\rightarrow 0}=\sigma^{\rm cs}_{\infty} \,. 
\end{align} 
This is not coincidental as these relations constitute special cases of a new duality between the corner coefficients of the two free CFTs: 
\beq
n^2\,\sigma_n^{\rm cs,\, f} = \sigma_{1/n}^{\rm f,\, cs}\,, \qquad \text{$n\!>\!0,\, n\in\mathbb R$.}
\labell{surprise}
\eeq
The duality interchanges the boson and fermion coefficients, together with $n\leftrightarrow 1/n$.
In particular $\sigma_n^{\rm cs}$ is fully determined by $\sigma_n^{\rm f}$, and vice versa. 
In appendix \ref{proof}, we prove this relation for any positive real \ren index, assuming 
our conjecture \req{conj2x}. As an immediate consequence of this duality, we can use our large-$n$ expansion eq.~\req{free-large-n}
to get the following expansion in the opposite limit: 
\begin{align}
\sigma_{n\rightarrow 0}^{\rm cs,\, f} = \frac{\sigma^{\rm f,\, cs}_\infty}{n^2}\ (1+n) \left[ 1+b_2^{\rm f,\, cs}\, n^2  
  - b_4^{\rm f,\, cs}\, n^4+\mathcal O\!\left( n^6\right) \right] \,,
\labell{small-n}
\end{align} 
with the same coefficients $b_k^{\rm f,\,cs}$ as in eq.~\reef{largenX}. Of course, combined with eq.~\req{cs0x}, these results imply that the coefficient $\sigma_\infty$ of one theory is determined by the thermal entropy coefficient $\cs$
of the ``dual theory''. 

This surprising duality between the corner coefficients in the two free theories has numerous other consequences. 
For example, we can rewrite it in terms of the conformal dimensions of the corresponding twist operators as
\beq
n\, h_n^{\rm cs,\, f}=-h_{1/n}^{\rm f,\, cs}\, . \labell{surprise2}
\eeq
We can go further
and use eqs.~\req{cs1} and \req{sifi} to establish relations between the thermal entropy coefficient of each field with the thermal energy density on 
the hyperbolic plane $\hyp{2}$ at $T_0$ of the other:\footnote{The form of this result relies on $\mathcal{E}^{\rm cs,\, f}(T\!=\!0)=0$ in our conventions.  In appendix \ref{proof}, we have adopted the conventions of \cite{igor5} in which $\mathcal{E}^{\rm cs,\, f}(T_0)=0$ instead. As explained below eq.~\reef{csinf} and in the appendix \ref{revolt}, the differences arise from the choice of the renormalization scheme for the partition function on $S^1\times\hyp{2}$.}
\beq
\mathcal{E}^{\rm cs,\, f}(T_0)=\frac{c_s^{\rm f ,\, cs}}{12\pi^3 R^3}\, , \labell{surprise3}
\eeq
where again $T_0=1/(2\pi R)$. In fact, combining eqs.~\reef{hn33} and \reef{surprise2}, we find a more general relation
between the energy densities of the two theories at generic temperatures on the hyperbolic cylinder,
\beq
\mathcal{E}^{\rm cs,\, f}(T_0^2/T)-\mathcal{E}^{\rm cs,\, f}(T_0) = -\frac{T^3}{T_0^3}\,\left(\mathcal{E}^{\rm f,\, cs}(T)-\mathcal{E}^{\rm f,\, cs}(T_0)\right)\,.
\labell{surprise33}
\eeq
In appendix \ref{proof}, we establish a general relation for the logarithm of the partition functions of the two theories on $S^1\times\hyp{2}$ and so there may be interesting relations, similar to eq.~\reef{surprise33}, for other thermal quantities.\footnote{There may also be interesting relations for the circle \ren entropies using the results in appendix \ref{circle} --- a topic we leave for future work.} These relations are relating the properties of one theory at temperature $T$ to the other at $T_0^2/T$. Hence,
in this regard, our duality is somewhat analogous to the Kramers--Wannier duality \cite{KW}, which relates quantities in the two-dimensional Ising model
at low and high temperatures.

We point out that a duality closer in spirit to Kramers--Wannier can be obtained by considering the free CFT containing both the complex scalar and the Dirac fermion. This CFT has $\mathcal N=2$ supersymmetry, as it consists of a
free chiral multiplet. Now, the corner function $a_n(\theta)$ is additive in that case as free fields factorize, 
implying $\sigma_n^{\ssc{\rm free}\,\mathcal N\!=2}\!=\sigma_n^{\rm cs} + \sigma_n^{\rm f}$.
The above Bose-Fermi duality \req{surprise} thus leads to  
\begin{align} 
 n^2\, \sigma_n^{\ssc{\rm free}\,\mathcal N\!=2} = \sigma_{1/n}^{\ssc{\rm free}\,\mathcal N\!=2} \,, \qquad n\!>\!0,\, n\in\mathbb R\,, 
\labell{susy}
\end{align}
which is a self-duality connecting $\sigma_q^{\ssc{\rm free}\,\mathcal N\!=2}$ at $q=n$ and $1/n$. In particular, in this theory, $\sigma_\infty$
acquires a concrete physical meaning as it is determined by the flat-space thermal entropy coefficient $\cs$ of the same theory.
It would be interesting to investigate the fate of this self-duality in the interacting $\mathcal N=2$ supersymmetric CFT, \ie the chiral Wess-Zumino model, which can be obtained by perturbing the theory of the free chiral multiplet.  

It is further natural to ask: Do similar entanglement dualities exist in dimensions other than $d=3$? Below we point out that this indeed occurs in $d=2$. 

\subsection{Entanglement self-duality in $d=2$}  

For any CFT in $d=2$, we have the following self-duality relation for the twist dimension:
\begin{align}
h^{\ssc(2)}_n=-h^{\ssc(2)}_{1/n}\, , 
\end{align}
as follows from eq.~\req{hn-twod}. This can then be rewritten in terms of the coefficient $\sigtwo_n$ of the logarithmic contribution to the \ren entropy of 
 a single interval as
\begin{align}
  n\, \sigma_n^{\rm\ssc (2)} = \sigma_{1/n}^{\rm \ssc (2)}\,,
\labell{2d-dual}
\end{align}
where we recall from eq.~\req{twod} that
\begin{align}
  \sigtwo_n = \frac{h^{\ssc(2)}_n}{n-1} = \frac{c}{12}\left(1+\frac{1}{n}\right)\, .
\end{align}
The relation \req{2d-dual} is clearly reminiscent of the $d=3$ Bose-Fermi duality presented above, and in particular 
of its manifestation as the self-duality in the free supersymmetric theory in eq.~\req{susy}.
As a special case of \req{2d-dual}, we have a Bose-Fermi duality between a free Dirac fermion and 
a real scalar (compact or not):
$n \,\sigma_n^{{\ssc (2)},\,\rm s}=\sigma_{1/n}^{{\ssc(2)},\,\rm f}$, as both have $c=1$.

Given the above results in $d=2,3$, we are led to propose that relations of the form $n^{d-1}\sigma^{\ssc(d)}_n\sim \sigma^{\ssc(d)}_{1/n}$ or $n^{d-2}h^{\ssc(d)}_n\sim -h^{\ssc(d)}_{1/n}$ 
might occur as well in other higher-dimensional theories. The dualities may connect $n$ to $1/n$ in two different theories, as in eq.~\reef{surprise}, or within the same
theory, as in eqs.~\req{susy} and \req{2d-dual}.

\section{Holography}
\labell{sec:holo}

Let us now consider strongly coupled holographic CFTs dual to Einstein gravity. We will use known results for the
twist dimension $h_n^{\rm hol}$ to obtain the $n$-dependence of the corner coefficient $\sigma_n^{\rm hol}$.
We begin with the AdS/CFT correspondence in its simplest setting, where it describes a given
$d$-dimensional boundary CFT in terms of $(d+1)$-dimensional gravity in the bulk with the action  
\begin{align}
I=\frac{1}{16\pi G}\int d^{d+1}x\sqrt{g}\left[\frac{d(d-1)}{L^2}+\mathcal R \right]\, ,
\labell{einst}
\end{align}
where $G$ is the $(d+1)$-dimensional Newton's constant, $L$ is the AdS$_{d+1}$ radius, and $\mathcal R$ is the Ricci scalar.  
In order to obtain $h_n^{\rm hol}$, we need to consider the thermal ensemble of the CFT on the hyperbolic geometry appearing in the construction of \cite{CHM}, which is then equivalent to a topological black hole \cite{topbh} with a hyperbolic horizon.
Using this connection together with eq.~\reef{hn3}, the scaling dimension of twist operators in the corresponding holographic CFTs was evaluated as \cite{holoren}
\begin{align} 
h^{\rm hol}_n=\frac{L^{d-1}}{8G}\, n \left(\frac{\sqrt{d^2 n^2-2 d\, n^2+1}+1}{d\, n}\right)^{d-2} \left(1-\frac{\left(\sqrt{d^2 n^2-2 d\,
   n^2+1}+1\right)^2}{d^2\, n^2}\right) \, .\labell{laugh}
\end{align}
In the following, we will focus our attention on the case of a three-dimensional boundary theory. Of course, one particular example of such a boundary CFT would be the supersymmetric gauge theory constructed in \cite{ABJM}, in an appropriate large $N$ and strong coupling limit.

Setting $d\!=\!3$ in eq.~\reef{laugh} and using the conjectured relation \req{conj2x}, we can predict the following form for the \ren
corner coefficient in these holographic CFTs:
\begin{align}\label{sighl}
\sigma^{\rm hol}_n=\frac{L^{2}}{108\pi G} \, \frac{(3n^2-2)\sqrt{1+3n^2}-2}{n^2(n-1)} \, .
\end{align}
Of course, at present, there is no method available by which we may evaluate these coefficients directly in a holographic framework. This expression can be easily used to compute the corner coefficient for arbitrary values of $n$, and some results are presented in Table \ref{tblh}. In the table, we normalized the corner coefficient by dividing by the central charge $\cthol=3L^2/(\pi^3G)$ \cite{biggb,graviton}. We recover the expected result for the corner coefficient in the entanglement entropy (\ie with $n=1$). 
That is, $\sigma_1^{\rm hol}/\cthol=\pi^2 /24$ \cite{bueno1,bueno2}. Let us note at this point that the numerical values for the $\sigma_n/\ctt$ in the holographic CFT are remarkably similar those for the free Dirac fermion, as shown in \rfig{sighCT}. The relative error is no more than $2.6\%$ for $n\ge1$ and $0.2$\% in the range $0\le n\le 1$. 
{\renewcommand{\arraystretch}{1.4}%
\begin{table}
  \centering
  \begin{tabular}{c||c|c|c|c|c|c|c|c} 
  $n $  &$n\to0$ &  1 & 2 & 3 & 4 & 5 & 6 & $n\to\infty$ \\
    \hline 
  $ \sigma^{\rm hol}_n /\cthol$  &$\frac{\pi^2}{81\,n^2}$ &  $\frac{\pi^2}{24 }$ & $\frac{\pi^2(5\sqrt{13}-1)}{648}$ & $\frac{\pi^2(25\sqrt{7}-1)}{2916}$ 
& $\frac{5\pi^2}{243}$ & $\frac{\pi^2(73\sqrt{19}-1)}{16200} $& $\frac{\pi^2(53\sqrt{109}-1)}{29160}$ &$\frac{\pi^2}{36\sqrt{3}}$
  \end{tabular} 
  \caption{Corner coefficient for various \reny indices in holographic CFTs dual to Einstein gravity. 
}
\label{tblh}   
\end{table}  
 
Turning now to the limits $n\to 0$ and $n\to 1$, we find
\begin{align}
\sigma^{\rm hol}_{n\to 0}&=\frac{L^{2}}{27\pi G}\left[\frac{1}{n^{2}}+\frac{1}{n}+\mathcal{O}\left(n^0\right)\right]\, ,  \notag\\
\sigma^{\rm hol}_n\big|_{n=1+\epsilon}&=\ \frac{L^{2}}{8\pi G}\,\left[1-\frac{7}{8}\epsilon+\mathcal{O}(\epsilon^2) \right]\, ,\labell{boop9}
\end{align} 
Comparing the $n\rightarrow 0$ limit above with eq.~\req{cs1}, we find
\beq
\cs^{\rm hol}=\frac{4\pi^2L^2}{9\,G}\, .
\eeq
This result exactly matches the thermal entropy coefficient for the holographic CFT calculated from a planar AdS$_4$ black hole in the bulk, \eg see \cite{bueno2}.

In the large-$n$ limit, we find:
\begin{align}\label{siho}
  \sigma_n^{\rm hol} = \sigma_\infty^{\rm hol} \, \left( 1+\frac{1}{n} \right) \left[ 1+\frac{b_2}{n^2} - \frac{b_4}{n^4} 
    + \mathcal O\!\left(\frac{1}{n^6}\right) \right] - \frac{2\sigma_\infty^{\rm hol}}{3\sqrt 3 n^3} \,,
\end{align} 
where 
\begin{align}\label{hooo}
b_2=\frac12\,,\quad b_4=\frac{2}{3 \sqrt{3}}-\frac{3}{8}\simeq 0.01\quad\;
{\rm and}\;\quad\sigma^{\rm hol}_{\infty}=\frac{L^2 }{12 \sqrt{3}\pi  G}=\frac{\pi^2}{36\sqrt{3}}\,\cthol\, .
\end{align}
We emphasize that we have shown analytically that the overall factor of $(1+1/n)$ appears in the first term of  eq.~\req{siho}. Of course, this was the structure conjectured for the large-$n$ expansions in eq.~\reef{free-large-n} for the free scalar and fermion CFTs. 
However in contrast to the free CFTs, the second term proportional to $1/n^3$ spoils this simple $n$-dependence for the holographic CFTs.
As explained in eq.~\reef{sifi}, the coefficient $\sigma_{\infty}$  is proportional to $\mathcal{E}(T_0)-\mathcal{E}(0)$, a difference of thermal energy densities on the hyperbolic cylinder. 
We note that the holographic calculations yield $\mathcal{E}(T_0)=0$ and $\mathcal{E}(0)=-L^2/(12\sqrt{3} \pi  G R^3)$
here \cite{holotherm}. These results contrast to those for the corresponding free field energy densities in section \ref{sec:free fields}, where $\mathcal{E}(0)$ vanishes while $\mathcal{E}(T_0)$ does not.

Let us close this section by mentioning that our results here should be easily generalizable to other holographic theories. In particular, it would be interesting to study these corner contributions for holographic theories whose bulk action contains higher-curvature terms, \eg using the results of \cite{holoren}.

\section{Corners with arbitrary opening angles} \label{sec:interp} 

In sections \ref{sec:free fields} and \ref{sec:holo}, we saw that the corner coefficient $\sigma_n$ has an interesting systematic dependence on the \ren index. Below, we argue that this dependence is a general property of the entire corner function $a_n(\theta)$, at all opening angles $\theta$. Further, in this section, we provide an efficient approach to approximating the angular dependence of the corner function $a_n(\theta)$, using knowledge of the coefficients $\kappa_n$ and $\sigma_n$, which control the asymptotic behaviour in the limits $\theta\to0$ and $\pi$, respectively.

\subsection{General $n$ dependence}

We now go beyond the nearly smooth limit and establish some general properties of the \ren index dependence of the corner function $a_n(\theta)$, valid for any opening angle $\theta$. Without loss of generality, we consider a region $V$  
containing a single corner with opening angle $0 <\theta <\pi$ but that is otherwise smooth,
as in \rfig{corner}. As it will be useful below, we recall that the \ren entropy associated with $V$ is then given by an expression of the form shown in eq.~\req{align}, \ie 
\beq
S_n(V)=\widetilde B_n\, \frac{{\cal A}(\partial V)}\delta - a_n(\theta)\, \log\left(\mathcal A(\partial V) /\delta\right)+\dotsb\,,
\labell{zoom}
\eeq
where we have written the leading term in a geometric fashion, with ${\cal A}(\partial V)$ being the total length of the entangling surface, and $\delta$ a UV cutoff.

Let us begin by considering the $n\to 0$ limit of the corner function. Here we begin by adapting the arguments that were encountered in section \ref{inside} for the present discussion. First, we step back to consider a circular region with an entirely smooth boundary. The problem of evaluating the \ren entropy for such a region in the vacuum of a general CFT can be mapped to one of examining a thermal ensemble on hyperbolic space $\hyp{2}$. However, it was observed by \cite{brian2} that in the $n\to 0$ limit the result is dominated by the high temperature behaviour, which in fact matches that in flat space. Hence we have that the \ren entropy for a circular entangling curve in three dimensions diverges as $S_n \sim 1/n^2$. In particular then, this argument indicates that as $n\to0$, the coefficient $\widetilde B_n$ in eq.~\reef{zoom} diverges as $1/n^2$
for a circular entangling surface. However, this coefficient is independent of the details of the geometry and so the same divergence appears in the leading area law contribution to the \ren entropy for any entangling surface. Now note that the corner makes a negative contribution to $S_n$ with the second term in eq.~\reef{zoom} but the total \ren entropy must always be positive. Hence we conclude that as $n\to0$, $a_n(\theta)$ can diverge at most as fast as $1/n^p$, with $p\leq 2$. Indeed, as $\theta\to \pi$, we have $a_n(\theta)\simeq \sigma_n\, (\theta-\pi)^2$ where $\sigma_n$ diverges 
precisely as $1/n^2$, eq.~\req{cs1}.  
Next, we can use the fact that $a_n(\theta)$ decreases monotonically for $0<\theta<\pi$, eq.~\req{monotonic}, to establish that for all angles $a_n(\theta)$ diverges at least 
as fast as $1/n^p$ with $p\geq 2$. Combining these two inequalities we are lead to $p=2$, \ie the corner
function diverges as $1/n^2$ for any $\theta$ as $n\to0$. 

Next we consider the opposite limit, $n\to\infty$. In this case, we know that $\sigma_n$ asymptotes to a finite value $\sigma_\infty>0$. 
Hence at this point, we again invoke the monotonically decreasing behaviour of the corner function to infer that $a_{n\to\infty}(\theta)$ must
either tend to a finite value or diverge, \ie it will behave as $n^q$ with $q\geq 0$ in the large-$n$ limit. However, if $a_{n\to\infty}(\theta)$
was growing (with $q>0$), the positivity of $S_n$ would require that the coefficient $\widetilde B_n$ of the area law term and hence the entire \ren entropy
are also growing with $n$ at least as fast as the corner function. However, it follows from considerations of information theory
that $S_n$ must satisfy various inequalities \cite{karol} and in particular, one finds that $\partial_n S_n\le 0$. That is, the \ren entropy
can not grow with increasing $n$ (at any point). Therefore, we are led to the conclusion
that the corner function remains finite as $n\to\infty$ (\ie $q=0$), again for any opening angle. 
Our calculations for the free CFTs confirm this result since the sharp corner coefficients $\kappa_n$ of the complex scalar
and the Dirac fermion approach finite values as $n\to\infty$, see section \ref{scl}.

The general asymptotics of the corner function found above for the limits $n\to0$ and $\infty$, as well as the behaviour for large and small angles, are schematically illustrated in \rfig{an-gen}.

\subsection{Simple approximations for $a_n(\theta)$}  
In this subsection we address the following question: For a given \ren index $n$, can one efficiently approximate the angle dependence 
of the corner function $a_n(\theta)$? We will show that by knowing the asymptotic coefficients $\kappa_n$ and $\sigma_n$ alone, 
one can construct a simple closed-form expression for $a_n(\theta)$ that is not only very accurate for all $\theta$, but also exact 
as $\theta\to 0,\pi$. 
We will use two physically motivated functions that obey the correct asymptotics as our building blocks:
\begin{align}\label{ato}
  \ato(\theta) &=  \frac{(\theta-\pi)^2}{\theta (2\pi-\theta)}\,, \\
  \att(\theta) &= 1+ (\pi-\theta)\cot\theta \,. 
\end{align}
The first one is the simplest algebraic function with a quadratic zero at $\theta=\pi$ and
with the pole at $\theta=0$, as $a_n(\theta)$ is expected to have on general grounds --- see eq.~\req{charg}. The additional factor of $(2\pi-\theta)$ in the denominator ensures the function satisfies the constraint $\ato(\theta)=\ato(2\pi-\theta)$, as required for the \ren entropy of pure states. 
Interestingly, precisely the same function yields the corner function $a_n(\theta)$ of certain Lifshitz quantum critical points \cite{shitz}.   
The second function $\att$ corresponds to the functional form of the corner contribution in the extensive mutual information model \cite{brian1,CH4,CH5}, as described above
at eq.~\req{emi}. Both functions satisfy the expected constraints in eq.~\reef{monotonic}. However, we note that they also satisfy a stronger constraint obtained for the entanglement entropy coefficient $a_1(\theta)$ using strong subadditivity and Lorentz invariance \cite{CHdir,geom}, \ie $\partial^2_\theta\tilde a^{(i)\!}(\theta)\ge |\partial_\theta\tilde a^{(i)\!}(\theta)|/\sin\theta$ for $0\leq\theta\le\pi$. Further, as mentioned in the introduction, reflection positivity also imposes a infinite tower of nonlinear constraints on the corner functions $a_n(\theta)$ \cite{positive1}. These constraints can be schematically written as $\det(\{\partial^{j+k+2}_{\theta} a_n(\theta)\}^{M-1}_{j,k=0} )\geq 0$, with $M\geq 1$
and $\theta\in[0,\pi]$. For example, with $M=1$, this reduces to the linear constraint $\partial_{\theta}^2 a_n\ge0$, but the second inequality with $M=2$ reads: 
\begin{align}
\partial_{\theta}^2 a_n\, \partial_{\theta}^4 a_n- (\partial^3_{\theta} a_n)^2 &\geq 0 \, .
\end{align}
We have managed to explicitly verify that our basis functions $\ato(\theta)$ and $\att(\theta)$ in eq.~\req{ato} satisfy these non-linear constraints for $M=2,\ldots,6$.   
\begin{figure}
\center
   \includegraphics[scale=.51]{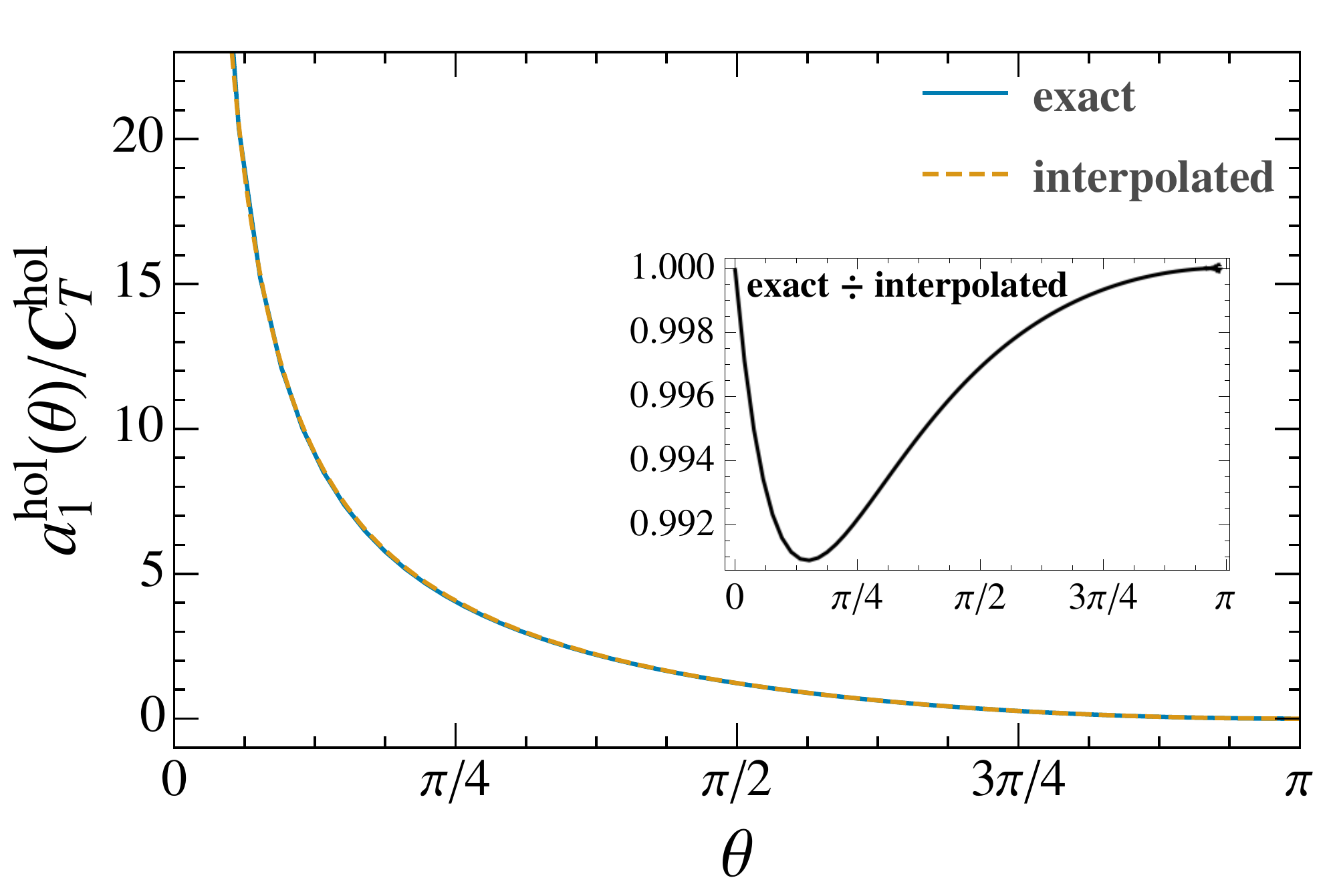}       
\caption{Holographic corner function $a^{\rm hol}_1(\theta)$ \cite{hirtak} normalized by $\cthol$, and the
approximate form obtained by interpolating the asymptotics with eq.~\req{interpolation}. The two lines can hardly be distinguished. The inset shows the ratio of the exact and extrapolated functions.} 
\labell{aCT}
\centering
\end{figure} 

We now construct an approximation $\tilde{a}_n(\theta)$ for the corner functions with a simple linear combination of $\ato$ and $\att$: 
\begin{align} \label{interpolation}
  a_n(\theta) \simeq \tilde a_n(\theta)\equiv  \lambda_n^{(1)} \ato(\theta) + \lambda_n^{(2)} \att(\theta)\,,  
\end{align}
where the constants $\lambda_n^{(i)}$ are fixed by requiring the right-hand side of eq.~\req{interpolation} to have the same asymptotics
as $a_n(\theta)$ in the limits $\theta\to0$ and $\pi$. Taylor expanding $\ato,\att$ in the these limits, we obtain:  
\begin{align}
  \theta\to0\ :&\ \ \ \frac{\pi}{2}\lambda^{(1)}_n + \pi \lambda^{(2)}_n  = \kappa_n\, , \labell{horse3}\\
  \theta\to\pi\ :&\ \ \ \frac{1}{\pi^2}\lambda^{(1)}_n + \frac{1}{3}\lambda^{(2)}_n = \sigma_n\, . \notag
\end{align}
The unique solution of eq.~\reef{horse3} for the two unknown coefficients is then:
\begin{align}
\lambda^{(1)}_n=2\pi\, \frac{\kappa_n-3 \pi  \sigma_n}{\pi ^2-6} \, , 
\qquad \lambda^{(2)}_n=-\frac{3}{\pi}\, \frac{2 \kappa_n - \pi^3 \sigma_n}{\pi^2-6} \, .  
\end{align} 
\begin{figure}
\centering
 \subfigure[]{\label{cs-interp}\includegraphics[scale=.43]{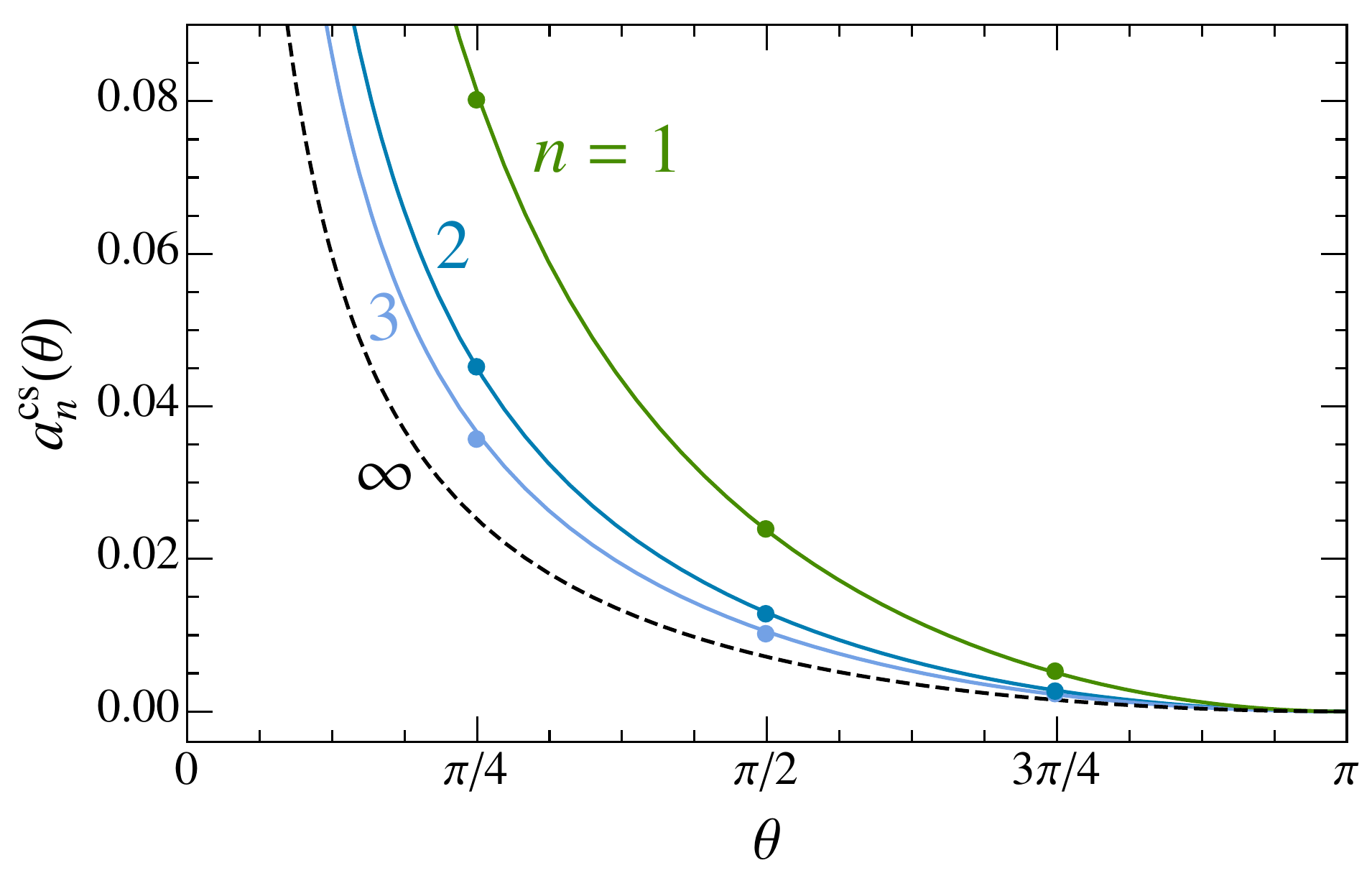}}  \hspace{.2cm}
 \subfigure[]{\label{f-interp} \includegraphics[scale=.43]{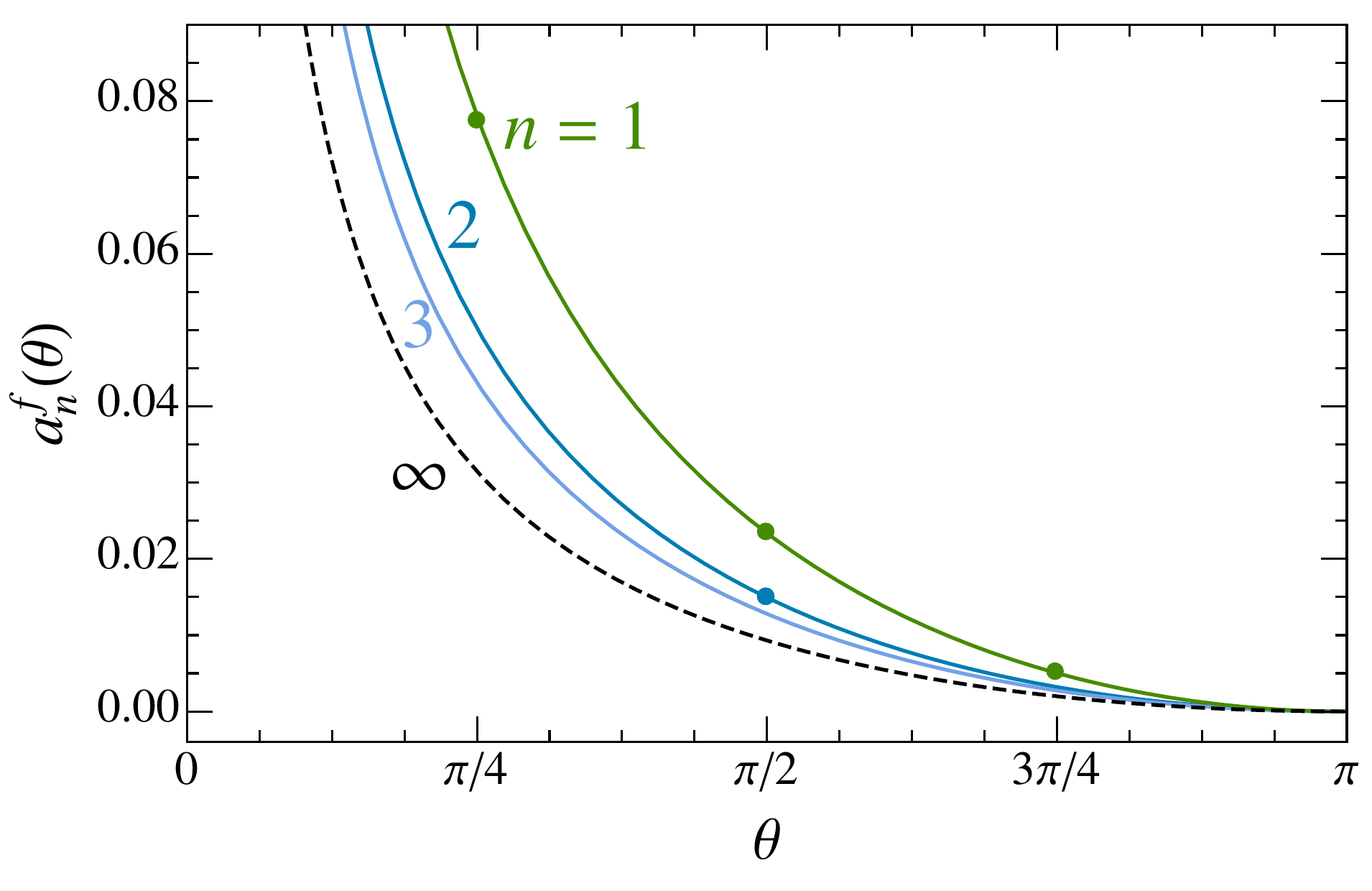}} 
\caption{\label{free-interp} Closed-form approximations for the corner function $a_n(\theta)$ at $n=1,2,3$
and $n=\infty$ for  (a) the free complex scalar and (b) the free Dirac fermion. The dots were obtained from numerical
lattice calculations \cite{CHdir}. The approximate functions obey the \emph{exact} asymptotics as $\theta\to 0,\pi$. } 
\end{figure} 

The holographic CFT dual to Einstein gravity provides a simple test of this approach \reef{interpolation}, at least for $n=1$. 
For this theory, we have an integral expression to evaluate $a_1^{\rm hol}(\theta)$ for any angle \cite{hirtak}, and the necessary 
coefficients are $\kappa_1^{\rm hol} =  \Gamma(3/4)^4\,L^2/(2\pi G)$ and $\sigma_1^{\rm hol}=L^2/(8\pi G)$, \eg see \cite{bueno2}. 
In \rfig{aCT}, we show the curves for the exact result for $a_1^{\rm hol}(\theta)/\cthol$ and the approximate function also normalized by $\cthol=3L^2/(\pi^3G)$. 
The agreement is seen to be excellent, with the relative error being less than 1\% over the entire range of angles.   

We can also use the values of $\sigma_n$ and $\kappa_n$ computed in the section \ref{gamble} to construct approximations of 
$a_n(\theta)$ for the free complex scalar and the free Dirac fermion. Now
in principle, the $a_{n}(\theta)$ functions can also be determined exactly for these free fields, as described in \cite{CHdir}. 
Unfortunately, their results are written implicitly in terms of several functions which must be determined through a complicated    
system of coupled non-linear algebraic and differential equations. We explicitly used this approach to test our approximation for $a_2^{\rm cs}(\theta)$ for the 
complex scalar, and we found that the difference between our approximation and the `exact' curve was comparable to our numerical errors in evaluating  $a_{n}(\theta)$ with the method of \cite{CHdir}. The numerical errors were of the order of 0.20\% while the maximum difference between the two curves was approximately 0.33\%. Clearly, these results suggest that our approach  
produces a faithful approximation to $a_n(\theta)$ for the free fields. In \rfig{free-interp}, we show our approximate corner functions for $n=1,2,3$ and $\infty$, both for the free scalar and fermion. The figure also includes the results of lattice simulations 
for $\theta=\pi/4,\ \pi/2$ and $3\pi/4$, in the cases where these are known,\footnote{We thank Horacio Casini for providing these lattice results, as well as his assistance 
in directly evaluating $a_2^{\rm cs}(\theta)$ using the method of \cite{CHdir}. We obtained $a_2^{\rm cs}(\theta)$ for $0.6 <\theta < 3.0$, but do not 
show it in \rfig{free-interp} since given the uncertainties described in the main text, the result would be indistinguishable from the approximate interpolation in the plot.}  
and again, we see that our interpolating curves agree very well with these lattice points. As an additional check, $a_4^{\rm cs}(\pi/2)$ was recently 
estimated to be $0.0086(2)$ by numerically studying the Goldstone phase of antiferromagnets on the square lattice \cite{Laflorencie}.
Using a different system, a numerical linked cluster expansion \cite{sahoo} yielded $0.0094(2)$. 
These values are in excellent agreement with our estimate $0.0094$, obtained with the simple interpolating method described in this section.  

Finally, we comment on the reasons behind the excellent accuracy of our interpolations. As noted in eq.~\req{monotonic}, $a_n(\theta)$   
is a monotonic convex function of $\theta$. Thus we may expect that once the behavior in the vicinity of the endpoints $\theta=0$ and $\pi$ is fixed (by our knowledge of $\kappa_n$
and $\sigma_n$), little freedom remains at intermediate angles. Our findings confirm this suspicion. Besides, we have checked that all the interpolating functions shown in Figures \ref{aCT} and \ref{free-interp} satisfy the first six inequalities arising from reflection positivity mentioned above. 


\section{Discussion}\label{disc}

In this paper, we have introduced a generalized version of the conjecture presented in \cite{bueno1,bueno2} for the coefficient appearing in the universal contribution in  the entanglement entropy for an almost smooth corner, 
\begin{align}\label{cog}
\sigma_1=\frac{\pi^2}{24}\,\ctt\, . 
\end{align}
In particular, we proposed that the analogous R\'enyi corner coefficients $\sigma_n$ --- see eq.~\req{charg} --- are related to the scaling dimension $h_n$ 
of the corresponding twist operators $\tau_n$ by
\begin{align}\label{conjji}
\sigma_n=\frac{1}{\pi}\, \frac{h_n}{n-1}\, .
\end{align}
This expression reduces to our previous conjecture in the limit $n\rightarrow 1$, and is therefore supported in this limit by the extensive results 
presented in \cite{bueno1,bueno2,rxm,prep2,Elvang}, corresponding to free scalars and fermions, as well as a general proof for holographic theories. Here we have checked the generalized version of the conjecture for a free conformally coupled complex scalar and a free massless Dirac fermion for integer \ren indices $1\leq n\leq 500$, as well as in the limit $n\rightarrow \infty$. The calculations involved in the evaluation of $\sigma_n$ and $h_n$ for the free fields are very different, both conceptually and computationally, which motivates us to conjecture that eq.~\req{conjji} is a universal expression for general three-dimensional CFTs. It would also 
be interesting to explore how these expressions get modified in non-conformal theories, 
see \eg \cite{shitz,Alishahiha:2015goa,Pang:2015lka}. 

Further, based on the properties of $h_n$ at small $n$, we have obtained that 
  \begin{align}
    \lim_{n\to0}\sigma_{n}= \frac{1}{12\pi^3}\, \frac{\cs}{n^2}\,,
  \end{align}
which yields a key observable, namely the thermal entropy density coefficient $\cs$ of the original CFT.  
As an interesting application of this result, we find that $\sigma_n$
for the O($N$) Wilson-Fisher IR fixed point in the $N\to\infty$ limit differs from its value
at the UV Gaussian fixed point. Indeed, the ratio of the entropy coefficients at $N\to\infty$ is
$\cs^{\rm IR}/\cs^{\rm UV}=4/5$ \cite{Sachdev93}. Hence $\sigma_n$ already distinguishes the entanglement structure of these two CFTs even in the absence  
of finite $N$ corrections, in contrast to the RG monotone $F$ \cite{Safdi11}.\footnote{Of course, the scaling dimension $h_n$ of the twist operator serves the same purpose.} 
This suggests that the full corner function $a_n(\theta)$ differs at other values of $n$ and $\theta$ between the $N\!\to\!\infty$ IR and UV fixed points.
It will be interesting to verify this claim.   

Our original motivation for the new conjecture \reef{conj2x} was as follows: Essentially, the corner coefficient $\sigma_n$ describes the universal response 
of the \ren entropy to a small deformation of the (originally smooth) entangling surface, as shown in \rfig{wide}. This small displacement of the twist 
operator can be accomplished by making appropriate insertions of the stress tensor, following the approach of \cite{vlad1} --- see also \cite{eric2,mark}. 
Hence in the case of the entanglement entropy, the response is determined by correlators of the stress tensor and the entanglement Hamiltonian \cite{vlad1,solos}. 
Since the latter is given by an integral of the stress tensor over the region $V$ \cite{CHM}, the calculation actually involves correlators of the 
stress tensor with itself and hence it is natural that the response is controlled by the central charge $\ctt$ as in eq.~\reef{conj1}. In the case of 
the \ren entropy (with $n\neq1$), the corresponding calculation instead involves correlators of the stress tensor and the twist operator and so this reasoning 
naturally suggests that the response would be determined by the scaling dimension $h_n$ as in eq.~\reef{conj2x}. This reasoning also outlines
a path potentially leading to a proof of our new conjecture, however, the detailed calculations are subtle and we must 
leave this topic for future work \cite{prep1}.

In eq.~\req{twod}, we observed that the universal contribution in the R\'enyi entropy of an interval in a two-dimensional CFT also contains a factor of $h_n/(n-1)$. 
This suggests that the form given in eq.~\reef{conjji} generalizes to higher dimension as well.  In particular, using the results of \cite{mark,sharp}, we have shown 
that the higher dimensional analogs of $\sigma_1$ describing the universal contribution in  the entanglement entropy for an almost smooth cone is proportional to 
the central charge $\ctt$ for general holographic CFTs in any number of dimensions \cite{prep2}.\footnote{Similar observations were recently made in \cite{rxm,Alishahiha:2015goa}.} Further, as shown in eq.~\reef{house}, $\partial_n h_n|_{n=1}$ is 
also related to $\ctt$ in general dimensions. Hence, it is natural to suggest that eq.~\req{conjji} will have a higher dimensional counterpart for the 
coefficients $\sigma^{\ssc(d)}_n$ controlling the response of the \ren entropy to an almost smooth conical defect in the entangling surface of the form  
\beq
\sigma^{\ssc(d)}_n = g(d)\,\frac{h^{\ssc(d)}_n}{n-1}\,,
\labell{house44}
\eeq
where $g(d)$ is a geometric factor that depends on the spacetime dimension, and $h^{\ssc(d)}_n$ is the scaling dimension of the corresponding twist operators in $d$ dimensions. Proving this conjecture would give a new perspective on the geometric character  of  \ren entropies in higher dimensions \cite{eric2,mark,ben}.

In section \ref{surprize}, we observed an intriguing duality between the corner coefficients of the free scalar and the free fermion. In particular, this relates the corner coefficients for inverse values of the R\'enyi index through the simple expression $n^2\sigma_{n}^{\rm cs,\, f}=\sigma_{1/n}^{\rm f,\, cs}$, which we proved in appendix \ref{proof}. Further implications of this duality include that the energy density of one theory on the hyperbolic cylinder at temperature $T$ is related to the corresponding energy density of the dual theory at $T_0^2/T$, as in eq.~\req{surprise33}. It would be interesting to explore  the physical implications of this new duality in more depth. Similarly, it would be worthwhile to search for similar relations for other theories in other spacetime dimensions.

Our proof of the Bose-Fermi duality makes use of the representation of $\log Z^{\rm cs,\, f}$ given in \cite{igor5} --- see appendices \ref{revolt} and \ref{proof}. This allows us to give the following alternative representation
 of $\sigma_n^{\rm cs\, ,f}$ 
 \beqa 
\sigma_n^{\rm cs}&=&\frac{n}{4\pi^2(n-1)} \left[\frac{3\zeta(3)}{4\pi^2} - \frac{2}{\pi^2}\int_0^{\infty}\!d\lambda\ 
\lambda^2\, \tanh \lambda\, \left(\coth (n \lambda)-1\right)\right]\,,
\labell{woow}\\
 \sigma_n^{\rm f}&=&\frac{n}{4\pi^2(n-1)} \left[\frac{\zeta(3)}{\pi^2} + \frac{2}{\pi^2}\int_0^{\infty}\!d\lambda\ 
\lambda^2\, \coth \lambda\, \left(\tanh (n \lambda)-1\right)\right] \,.
\nonumber
\eeqa
It would be interesting prove the equivalence of these integral expressions with those in eq.~\reef{confs}, which were used in the main text. Alternatively, it also appears that these expressions may be somewhat simpler to relate to the sums in eq.~\req{sumsig}, than those in eq.~\req{confs}.

In this section \ref{sec:interp}, we devised an approach to constructing a simple analytic function which gives a (very) good approximation to the entire corner function $a_n(\theta)$ for all angles. These functions make an informed interpolation between the asymptotic limits \reef{charg} at $\theta\to0$ and $\pi$. Hence the only data necessary to construct these approximations is the almost smooth corner coefficient $\sigma_n$ and the small angle coefficient $\kappa_n$ in eq.~\reef{charg}. Of course, our conjecture \req{conjji} indicates that the coefficient $\sigma_n$ can be accessed by a different calculation. In particular, the scaling dimension $h_n$ is determined by the leading singularity in correlation function of the stress tensor with a twist operator, as shown in eq.~\reef{sing2}. Similarly, the small angle coefficient $\kappa_n$ can be accessed by an alternative calculation. In particular, as described in \cite{CHdir,bueno2}, the same coefficient controls the universal contribution in the \ren entropy of an narrow strip. It would be interesting to investigate situations where these two alternate calculations would allow the corner contribution to be determined more simply than by evaluating the \ren entropy for entangling surfaces with corners of varying angles.\\
 
\textbf{\textit{Note added}} --- After the first version of this paper appeared, Dowker \cite{Dowker:2015qta} 
presented an analytic proof of our conjecture \req{conjji} in the special case of a free scalar using eq.~\reef{sumsig} and his own results for the scaling dimensions $h_n^{\rm cs}$. \\

\section*{Acknowledgments} 

\noindent We thank Horacio Casini for many valuable explanations and comments, particularly regarding the free field calculations. We are also thankful to Lorenzo Bianchi, John Cardy,
Aitor Lewkowycz, Patrick Meessen, Marco Meineri, Roger Melko, Joel Moore, Subir Sachdev, 
Misha Smolkin, and Miles Stoudenmire for useful discussions and comments. Research at Perimeter Institute is supported by the 
Government of Canada through Industry Canada and by the Province of Ontario through the Ministry of Research \& Innovation. The work of 
PB has been supported by the JAE-predoc grant JAEPre 2011 00452.  
RCM acknowledges support from an NSERC Discovery grant and funding from the Canadian Institute for Advanced Research.

\appendix
\section{Evaluation of $h_n$ for the free CFTs}
\label{hn_sum}
In this appendix, we evaluate the integral expressions in eq.~\reef{confw} for the scaling dimension of the twist operator with odd \ren index $n>1$, in the free conformal field theories. Using the same integrals along with eq.~\reef{conj2x}, we also evaluate $\sigma_n$ in closed-form for various rational \emph{non-integer} values of $n$. These results 
provide an interesting explicit check of the Bose-Fermi duality discussed in section \ref{surprize}, and proven in appendix \ref{proof}.  

\subsection{Complex scalar}
We begin with the conformal dimension $h_n$ in the theory of a free complex scalar, which may be written as:
\begin{align}
h_n^{\rm cs}&=\frac{1}{8\pi n^2}\int_{-\infty}^{\infty}\frac{du}{\sinh u}\left[\frac{\cosh(u/n)}{\sinh^3(u/n)}-n^3\,\frac{\cosh u}{\sinh^3u} \right]\, .
 \labell{confw2}
\end{align}
Here we have used the fact that the integrand is even to extend the range of integration over the entire real line. In passing, we also note that the integrand is finite at the origin, where it becomes $(n^4-1)/(15n)+O(u^2)$. Further we note that the integrand decays exponentially fast as $u\to\pm\infty$. Next we observe that if we shift the contour in the complex $u$ plane from the real axis to $\Im(u)= i\pi n$, then the integrand is invariant up to an overall phase of $e^{i n\pi}=(-)^n$.  Hence for odd $n$, we may re-express the conformal weight in terms of complex contour integral which runs along the contour $C$, illustrated in \rfig{contour}:
\beq
h_n^{\rm cs}=\frac{1}{16\pi n^2}\oint_C\frac{du}{\sinh u}\left[\frac{\cosh(u/n)}{\sinh^3(u/n)}-n^3\,\frac{\cosh u}{\sinh^3u} \right]\
\, .\labell{confw3}
\eeq
\begin{figure}
\center
   \includegraphics[scale=.5]{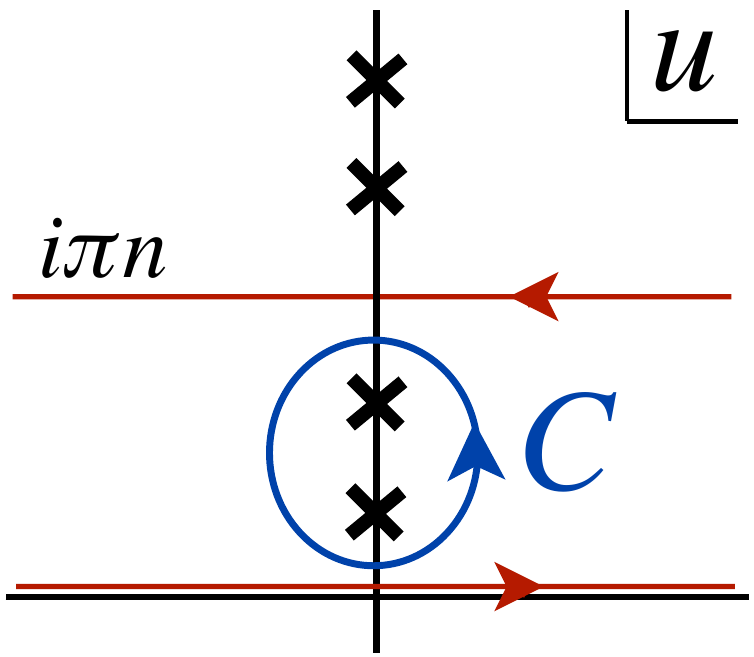}
\caption{ The integration contour $C$ in the complex $u$ plane. Poles
occur at $u=i\pi k$, for all integers $k$ that are not multiples of $n$. Here we illustrate the case $n=3$, with
two poles lying inside $C$. }\labell{contour}
\centering
\end{figure}  
\noindent Thus the result will be expressed in terms of the residues of the poles found to lie within 
the contour $C$. The integrand within the contour is only singular along the imaginary axis 
where $\sinh u$ vanishes, 
\ie at $u= i \pi k$ with $k=1,2,\cdots,n-1$. The second term in the integrand actually has a quartic and quadratic pole at each of these points but neither of these contribute to the closed contour integral. Hence only the first term contributes with the result
\beq
h_n^{\rm cs}=\frac{1}{8n^2}\sum_{k=1}^{n-1}(-)^{k+1}\frac{\cos(\pi k/n)}{\sin^3(\pi k/n)}
\, .\labell{confw4}
\eeq
We note again that this result only applies for odd values of $n$. One can confirm that this sum vanishes for $n$ even, as expected. We have also explicitly verified that the above expression reproduces the value of $h^{\rm cs}_n$ given by directly evaluating the integral in eq.~\reef{confw} for odd values of $n$ up to $n=500$.  
Combined with the new conjecture \req{conj2x}, the above expression yields the following sum for the corresponding corner coefficient  (for odd $n>1$):
\beq
\sigs_n=\frac{1}{8\pi n^2(n-1)}\sum_{k=1}^{n-1}(-)^{k+1}\frac{\cos(\pi k/n)}{\sin^3(\pi k/n)}
\, .\labell{sigw4}
\eeq
Unfortunately, the sum appearing here does not seem to be simply related to that appearing in eq.~\req{sumsig}.

\subsection{Dirac fermion}
Now we consider the conformal dimension of the twist operator in the theory of a free Dirac fermion. Again the integrand in eq.~\req{confw} is even and so we extend the integral to the entire real line:
\begin{align}
h_n^{\rm f}&=\frac{1}{16\pi n^2}\int_{-\infty}^{\infty}\frac{du}{\tanh u}\left[n^3\,\frac{2+\sinh^2 u}{\sinh^3u}
-\frac{2+\sinh^2 (u/n)}{\sinh^3(u/n)} \right]\, . \labell{confw2a}
\end{align}
Again, we note that the integrand is finite at the origin, where it becomes $7(n^4-1)/(30n)+O(u^2)$. Further we note that the integrand decays exponentially fast as $u\to\pm\infty$. Next we observe that if we shift the contour in the complex $u$ plane from the real axis to $\Im(u)= i\pi n$, then the integrand is invariant up to an overall change of sign, but only for odd $n$. Hence we may re-express the conformal weight in terms of complex contour integral which 
runs along the contour $C$, illustrated in \rfig{contour}:
\beq
h_n^{\rm f}=\frac{1}{32\pi n^2}\oint_C\frac{du}{\tanh u}\left[n^3\,\frac{2+\sinh^2 u}{\sinh^3u}
-\frac{2+\sinh^2 (u/n)}{\sinh^3(u/n)} \right]\, . \labell{confw3a}
\eeq
Hence the result can again be expressed in terms of the residues of the poles found to lie within the contour $C$. 
The integrand in $C$ is only singular along the imaginary axis where $\sinh(u)$ vanishes, \ie at $u=i \pi k$ with 
$k=1,2,\cdots,n-1$. Again, the first term in the integrand actually has a quartic and quadratic pole at 
each of these points and neither of these contribute to the closed contour integral. Hence only the second 
term contributes with the result
\beq
h_n^{\rm f}
=\frac{1}{16n^2}\sum_{k=1}^{n-1}\left[\frac{2}{\sin^3(\pi k/n)}-\frac{1}{\sin(\pi k/n)}\right]
\, .\labell{confw4a}
\eeq
Again, this result only applies for $n$ odd. In this case, the sum does not vanish for $n$ even, but the result should not be taken as the correct value of $h_n^{\rm f}$. We have again explicitly verified that this 
sum reproduces the value of $h^{\rm f}_n$ found by directly evaluating the integral in eq.~\reef{confw} for odd values of $n$ up to $n=500$. Combining the above sum with the new conjecture \req{conj2x}, we find the following expression for the corresponding corner coefficient (for odd $n>1$):
\beq
\sigf_n=\frac{1}{16\pi n^2(n-1)}\sum_{k=1}^{n-1}\left[\frac{2}{\sin^3(\pi k/n)}-\frac{1}{\sin(\pi k/n)}\right]
\, .\labell{sigw4a}
\eeq
Here again, this new sum does not seem to be simply related to the sum in eq.~\req{sumsig} for the same coefficients.

\subsection{Rational values of $n$}

Using our integrals \reef{confw} for the scaling dimensions $h_n$ of the free scalar and fermion, we can evaluate
$\sigma_n$ in closed-form at various rational \emph{non-integer} values of $n$. We list select
results in Table \ref{tbl-rational}. Comparing the results for $n=2/3$ and $3/2$, we see that they explicitly satisfy the Bose-Fermi duality $n^2\sigma^{\rm sc}_n=\sigma_{1/n}^{\rm f}$ in eq.~\reef{surprise}.
Similarly, comparing $n=1/6$, $1/5$, $1/4$, $1/3$ and $1/2$ with the dual results in Table \ref{tbl2}, we see that they obey the same duality relation.
We provide a general proof of the duality eq.~\reef{surprise} in appendix \ref{proof}. 
\begin{table}   
  \centering
  \begin{tabular}{c||c|c|c|c|c|c|c}  
  $n$ & $1/6$ & $1/5$ & $1/4$ & $1/3$ & $2/3$  & $1/2$ & $3/2$\\
    \hline\hline \rule{0pt}{1.5em}
  $\sigs_n$ & $\frac{261+20 \sqrt{3}}{720 \pi }$ & $\frac{\sqrt{425+58\sqrt 5}}{80\pi}$ & $\frac{1+6\sqrt 2}{48\pi}$ & $\frac{5}{24\pi\sqrt 3}$ 
& $\frac{20 \sqrt{3} \pi -81}{216 \pi ^2}$ & $\frac{1}{16\pi}$ & $\frac{243-128 \sqrt{3}}{1296 \pi }$ 
 \\ \hline \rule{0pt}{1.5em}
   $\sigf_n$ & $\frac{81+34 \sqrt{3} \pi }{270 \pi ^2}$  & $\frac{\sqrt{25-2\sqrt 5}}{20\pi}$ & $\frac{8+3\pi}{36\pi^2}$ & $\frac{1}{6\pi\sqrt 3}$ & 
$\frac{243-128 \sqrt{3}}{576 \pi }$ & $\frac{1}{6\pi^2}$ & $\frac{20 \sqrt{3} \pi -81}{486\pi^2}$ 
  \end{tabular}  
  \caption{Corner coefficient $\sigma_n$ for various rational values of the \reny index for the complex scalar and Dirac CFTs in $d=3$.} 
\labell{tbl-rational}   
\end{table}

\section{Sharp corner coefficient calculation}
\label{kappas}
In section \ref{scl}, we computed the sharp corner coefficients $\kappa_n$, $n=2,3,4$ and $n=\infty$ 
for the complex scalar and the Dirac fermion. Here we explain how these calculations are performed following the approach of \cite{CHdir}. In particular, we exploit the relation between these coefficients and the so-called `entropic c-functions' for a family of corresponding massive free fields in $d=2$. As was noted in section \ref{scl}, the $\kappa^{\rm cs,\, f}_n$ coefficients are given by
\begin{align}
\kappa_n^{\rm cs,\,  f}=\int_0^{\infty}\!\frac{dt}{\pi}\,  c^{\rm cs,\,  f}_n(t)\, , \qquad \text{where}\qquad
c^{\rm cs,\,  f}_n(m\ell)=\ell\,\frac{dS_n^{\rm cs,\,  f}(m\ell)}{d\ell}\,,
\end{align}
where $S_n^{\rm cs,\,  f}(m\ell)$ are the \ren entropies of $d=2$ free fields with mass $m$ for a single interval of length $\ell$.
The $c_n(t)$ for the complex scalar and the Dirac fermion can be evaluated as 
\begin{align}
c_{n}^{ \rm cs}=\frac{1}{1-n}\sum_{k=1}^{n-1}\omega^{ \rm cs}_{k/n}(t)\, , \qquad c_{n}^{ \rm f}=\frac{2}{1-n}\sum_{k>0}^{(n-1)/2}\omega^{ \rm f}_{k/n}(t)\, ,
\end{align}
respectively, where the $\omega^{ \rm cs,\,f}_{a}$ are given by
\begin{align}
\omega_a(t)=-\int_t^{\infty}dy\,y\, u^2_{a}(y)\, ,\labell{quack}
\end{align}
and the functions $u_a(y)$ satisfy, respectively, the differential equations
\begin{align}\label{eqs}
u_{a,\,  \rm cs}^{\prime\prime}(t)+\frac{u^{\prime}_{a,\,  \rm cs}(t)}{t}-\frac{u_{a,\,  \rm cs}(t)}{1+u_{a,\,  \rm cs}^2(t)}\,{u^{\prime}}_{a,\,  \rm cs}^2(t)-u_{a,\,  \rm sc}(t)(1+u_{a,\,  \rm cs}^2(t))-\frac{(2a-1)^2}{t^2}\,\frac{u_{a,\,  \rm cs}(t)}{1+u_{a,\,  \rm cs}^2(t)}=0\, ,
\end{align}
\begin{align}\label{bds}
u_{a,\,  \rm cs}(t\rightarrow 0)&=\frac{-1}{t(\log(t/2)+2\gamma_{\rm E}+(\psi[a]+\psi[1-a])/2)}\, ,\\ \notag
u_{a,\,  \rm cs}(t\rightarrow \infty)&=\frac{2}{\pi}\sin(a\pi)K_{1-2a}(t)\, ,
\end{align}
and 
\begin{align}\label{eqf}
u^{\prime\prime}_{a,\,  \rm f}(t)+\frac{u^{\prime}_{a,\,  \rm f}(t)}{t}+\frac{u_{a,\,  \rm f}(t)}{1-u_{a,\,  \rm f}^2(t)}\,{u^{\prime}}_{a,\,  \rm f}^2(t)-u_{a,\,  \rm f}(t)(1-u_{a,\,  \rm f}^2(t))-\frac{4a^2}{t^2}\,\frac{u_{a,\,  \rm f}(t)}{1-u_{a,\,  \rm f}^2(t)}=0\, ,
\end{align}
\begin{align}\label{bdf}
u_{a,\,  \rm f}(t\rightarrow 0)&=-2a(\log(t/2)+2\gamma_{\rm E}+(\psi[a]+\psi[-a])/2)\, , \\ \notag 
u_{a,\,  \rm f}(t\rightarrow \infty)&=\frac{2}{\pi}\sin(a\pi)K_{2a}(t)\, .
\end{align}
The differential equations can also be used to eliminate the integral in eq.~\reef{quack} producing
\begin{align}\labell{simpom2}  
\omega^{\rm cs}_a(t)&
=-  \frac{t^2{u^{\prime}}_{a,\,  \rm cs}^2-{u}_{a,\,  \rm cs}^2((1-2a)^2+t^2(1+{u}_{a,\,  \rm cs}^2))}{2(1+{u}_{a,\,  \rm cs}^2)} \, ,\\ \labell{simpom3} 
\omega^{\rm f}_a(t)&
=  -\frac{t^2{u^{\prime}}_{a,\,  \rm f}^2-{u}_{a,\,  \rm f}^2(4a^2+t^2(1-{u}_{a,\,  \rm f}^2))}{2(1-{u}_{a,\, \rm f}^2)}\, .
\end{align}
The final expressions for the $\kappa_n^{\rm cs,\, f}$ are then given in eq.~\req{kas}. Hence, our approach is to numerically solve eqs.~\req{eqs} and \req{eqf} subject to the boundary conditions \req{bds} and \req{bdf}, respectively, and then use the results to evaluate eq.~\req{kas}.

Let us add here that we can evaluate $\kappa^{\rm cs,\,f}$ corresponding to the special case $n=1$ with a similar approach using two-dimensional entropic c-functions. In particular, we have
\beq
\kappa^{\rm cs,\,f}= \int_0^\infty \frac{dt}{\pi}\ c^{\rm cs,\,f}(t) \, ,
\labell{none}
\eeq
with
\begin{align}\labell{ccsf1}
c^{\rm cs}(t)&=-\pi\int_0^\infty \frac{db}{\cosh^2(\pi b)}\ \omega^{\rm cs}_{1/2-ib}(t)\,, \\ \labell{ccsf2}
c^{\rm f}(t)&=2\pi\int_0^\infty \frac{db}{\sinh^2(\pi b)}\ \omega^{\rm f}_{-ib}(t)\, .
\end{align}
The values of these coefficients were evaluated by \cite{CHdir} and appear in Table~\ref{tbl3} under $n=1$.


\section{Relation between $\sigma_{ \infty}$ and $\mathcal{F}_{\infty}$}
\labell{revolt}

In comparing the behaviour of eq.~\reef{sumsig} for large $n$, the authors of \cite{Elvang} observed a relation \reef{helv} between $\sigma_{ \infty}$ for the free fields and corresponding results for the logarithm of the partition function on the hyperbolic cylinder $S^1\times \mathbb{H}^2$, given in \cite{igor5}.
We will provide an explanation of this relation in this appendix and at the same time, we show that the same result holds for any three-dimensional CFT. 

Following \cite{igor5}, we begin by defining the logarithm of the thermal partition function evaluated at $T=T_0/n$ 
\beq\labell{parr}
\hF_n \equiv -\log Z(T_0/n)\,,
\eeq
which is now readily connected to the corner coefficient $\sigma_n$ through eq.~\reef{sin33}. In particular, it is straightforward to show that eq.~\reef{ener} for the desired energy densities can be re-expressed as
\begin{align}
 \mathcal{E}(T_0/n)= \frac{1}{2\pi R\,V_{\hyp{2}}}\,\partial_n\hF_n\,,
 \labell{parr1} 
\end{align}
and hence eq.~\reef{sin33} can be written as
\beq\labell{parr2}
\sigma_n=-\frac{R^2}{2\pi\, V_{\hyp{2}}}\,\frac{n}{n-1}\,\left(\partial_n\hF_n-\partial_n\hF_n\big|_{n=1} \right)
\eeq 
Now, of course, the quantity $\hF_n$ is infrared divergent because it involves an integration over the infinite volume 
of the hyperbolic plane $\hyp{2}$. However, both of the previous two formulae compensate for this divergence by dividing
by $V_{\hyp{2}}$ to produce a finite quantity. One can produce expressions involving only finite quantities by focusing on 
the universal contribution to $\hF_n$, which amounts here to only retaining the finite regulator-independent contribution. The IR regulator also renders the volume finite and one only keeps the regulator-independent term in $V_{\hyp{2}}$. 
In particular, examining eq.~\reef{regulation}, we have $V^{\rm univ}_{\hyp{2}}=-2\pi R^2$. 
Hence eq.~\reef{parr2} can be written as
\beq\labell{parr3}
\sigma_n=\frac{1}{4\pi^2}\,\frac{n}{n-1}\,\left(\partial^{\vphantom i}_n\hF^{\rm univ}_n-\partial^{\vphantom i}_n\hF^{\rm univ}_n\big|_{n=1} \right)\,.
\eeq

However, before proceeding, we must first consider a difference in the conventions in the present paper and \cite{igor5}, which was alluded to in the discussion after eq.~\reef{csinf}. With the heat kernel regularization used to produce the partition functions \reef{part2} for the free fields \cite{twist}, one finds a nonvanishing energy density at $T=T_0$. On the other hand, with the approach used in \cite{igor5}, $\mathcal{E}(T_0)=0$ and hence  $\partial_n^{\vphantom i}\hF^{\rm univ}_n\big|_{n=1}=0$ as a consequence of eq.~\reef{parr1}. Since our present goal is to explain the result in eq.~\reef{helv}, we adopt the latter convention for the moment. Of course, this produces a simplification in eq.~\reef{parr3} and we are left with the following expression for the corner coefficient,
\beq\labell{parr4}
\sigma_n=\frac{1}{4\pi^2}\,\frac{n}{n-1}\,\partial_n^{\vphantom i}\hF^{\rm univ}_n\,.
\eeq
Finally, we consider the limit $n\to\infty$ for which one finds a linear growth in $\hF^{\rm univ}_n$. Hence defining $\mathcal{F}_\infty$ as the coefficient of the leading term, \ie $\hF^{\rm univ}_n\sim n\,\mathcal{F}_\infty$, the above expression reproduces the desired relation \reef{helv}
\begin{align}\labell{conjeX}
 \sigma_{\infty}=\frac{1 }{4\pi^2}\,\mathcal{F}_{\infty}\, ,
\end{align}
which was observed in \cite{Elvang} for the free CFTs. Of course, our analysis here made no reference to the free field theories discussed there and so we have established that eq.~\reef{helv} holds generally for any three-dimensional CFT
--- assuming that the renormalization scheme is chosen such that $\mathcal{E}(T_0)=0$.  

To close, let us return to the difference in the conventions adopted here and in \cite{igor5}. As noted above and discussed below eq.~\reef{csinf}, differences in the renormalization of the Euclidean partition function produce a shift in $\hF_n$:
\beq
\left[\hF_n\right]_{\rm here}= \left[\hF_n - n \ \big(\partial_n \hF_n\big)_{\!n=\infty}\right]_{\cite{igor5}}\ .
\labell{def787}
\eeq
Hence let us consider the form of eq.~\reef{helv} if we instead adopt the conventions of the present paper. First, we note that with eq.~\reef{def787},  $\partial_n^{\vphantom i}\hF^{\rm univ}_n$ vanishes in the limit $n\to\infty$. Therefore eq.~\reef{parr3} reduces to the following expression
\beq\labell{conje3}
\sigma_{\infty}=-\frac{1}{4\pi^2}\,\partial_n^{\vphantom i}\hF^{\rm univ}_n\big|_{n=1} = R^3\,\mathcal{E}(T_0)\,.
\eeq
Again, we emphasize that although eqs.~\reef{conjeX} and \reef{conje3} are dissimilar in appearance, they both yield the same numerical value for $\sigma_\infty$.


\section{Proof of the Bose-Fermi duality}\labell{proof}

In this appendix, we prove the intriguing duality relation \req{surprise} relating the \ren corner coefficients of the free scalar and 
the free Dirac fermion through
\begin{align}\label{surprisea} 
n^2\,\sigs_{n}=\sigf_{1/n}\,, \qquad \text{for $n>0$, $n\in\mathbb R$.}
\end{align}
The proof is most simply presented by adopting the conventions and notation of \cite{igor5}, as described in the previous appendix. In this case, we can use eq.~\reef{parr4}  to rewrite the duality relation \req{surprisea} as
\begin{align}\labell{surpri}
n^3 \partial^{\vphantom d}_{n}\hF_n^{\rm univ,\,  cs}+\left. \partial_{q}^{\vphantom d}\hF_q^{\rm univ,\, f}\right\vert_{q= 1/n} =0\, .
\end{align}
Now, ref.~\cite{igor5} provides the following simple expressions for $\hF_n^{\rm univ,\,  cs}$ and $\hF_n^{\rm univ,\,  f}$ 
\begin{align}
\hF_n^{\rm univ,\,  cs} &= n\frac{3\zeta(3)}{4\pi^2} -\frac{2}{\pi^2} \int_0^\infty\! d\lambda\, \lambda \tanh(\lambda)\log\left(1-e^{-2 n \lambda} \right) 
\,, \\
\hF_n^{\rm univ,\,  f} &= n\frac{\zeta(3)}{\pi^2} + \frac{2}{\pi^2} \int_0^\infty\! d\lambda\,  \lambda \coth(\lambda)\log\left(1+e^{-2 n \lambda} \right)  \,.
\end{align}
Taking derivatives with respect to $n$ of the expressions above then yields
\begin{align}
\partial _n^{\vphantom d}\hF_n^{\rm univ,\,  cs} &= \frac{3\zeta(3)}{4\pi^2} -\frac{2}{\pi^2} \int_0^\infty\! d\lambda\, \lambda^2 \tanh(\lambda)\left(\coth(n\lambda)-1 \right) \,, \\ 
\partial^{\vphantom d}_n \hF_n^{\rm univ,\,  f} &= \frac{\zeta(3)}{\pi^2} + \frac{2}{\pi^2} \int_0^{\infty}\! d\lambda\,  \lambda^2 \coth(\lambda)
\left(\tanh(n\lambda)-1 \right)  \, . 
\end{align}
We note in passing that these can be used to produce yet another integral representation of $\sigma_n^{\rm cs\,, f}$ through eq.~\req{parr4} --- see eq.~\req{woow}. 
Now, after some manipulations, including a change of variables $\lambda \rightarrow \lambda/n$ in the scalar integral, we find
\begin{align}
n^3 \partial^{\vphantom d}_n\hF_n^{\rm univ,\,  cs} &= \frac{3\zeta(3)}{4\pi^2}n^3 -\frac{2}{\pi^2} \int_0^\infty \!d\lambda\, \lambda^2 \tanh(\lambda/n)\left(\coth(\lambda)-1 \right)  
\,, \\
\left.\partial^{\vphantom d}_q \hF_q^{\rm univ,\,  f}\right\vert_{q= 1/n} &= \frac{\zeta(3)}{\pi^2}+ \frac{2}{\pi^2} \int_0^{\infty} d\lambda\,  \lambda^2 \coth(\lambda)(\tanh(\lambda/n)-1 )  \, .
\end{align}
Hence the right-hand side of eq.~\req{surpri} becomes
\begin{align} \labell{idd}
n^3 \partial^{\vphantom d}_{n}\hF_n^{\rm univ,\,  cs}+\left.\partial^{\vphantom d}_q \hF_q^{\rm univ,\,  f}\right\vert_{q= 1/n}= 
\frac{\zeta(3)}{4\pi^2}(3n^3+4) + 
\frac{2}{\pi^2} \int_0^{\infty}\!d\lambda\, \lambda^2 \left[\tanh(\lambda /n)-\coth\lambda \right] \, .
\end{align}
Finally, this expression vanishes identically by virtue of the exact result for the integral
\begin{align}
 \int_0^{\infty} \!d\lambda\, \lambda^2 \left[\tanh(\lambda /n)-\coth \lambda \right]=- \frac{\zeta(3)}{8}(3n^3+4)\,, \quad \forall \, \, n > 0\, .
\end{align}
This completes the proof of eq.~\reef{surpri}, and hence of the duality eq.~\req{surprisea}. 

If we apply the simple identity $\partial_qf(q)|_{q=1/n}=-n^2\partial_nf(1/n)$, we can express eq.~\reef{surpri} as
\begin{align}\labell{surpri2}
 \partial_{n}^{\vphantom d}\hF_{1/n}^{\rm univ,\, f}=n\, \partial^{\vphantom d}_{n}\hF_n^{\rm univ,\,  cs}\, ,
\end{align}
which can be integrated to yield
\begin{align}\labell{surpri3}
 \hF_{1/n}^{\rm univ,\, f}=n\, \hF_n^{\rm univ,\,  cs}-\int_1^n\!\!dq\, \hF_q^{\rm univ,\,  cs}-\left(\hF_1^{\rm univ,\,  cs}-\hF_{1}^{\rm univ,\, f}\right)\, .
\end{align}
Of course, $\hF_{1}^{\rm univ}$ is equivalent to the universal coefficient in the sphere partition function which plays a role in the $F$-theorem \cite{Safdi11,cthem}.
Eq.~\reef{surpri3} shows that the thermal partition functions of the two free theories are not simply exchanged under $n\leftrightarrow 1/n$, but they are still related in a relatively simple manner. 


\section{Relation to circular region  R\'enyi entropies} \labell{circle}

Let us first recall eq.~\reef{d3h}:
\begin{align}
\left.\partial_n h_n\right\vert_{n=1}=2\pi^{\frac{d}{2}+1}\frac{\Gamma(d/2)}{\Gamma(d+2)}\ctt\, .
\end{align}
This relation was first observed to hold for certain holographic theories in \cite{holoren} and later proven for general CFTs in \cite{twist}.
This result was also connected to a similar relation found in \cite{eric1} for derivatives of the R\'enyi entropy $S_n$ of a spherical entangling region 
\begin{align}
\left.\partial_n S_n\right\vert_{n=1}=-\pi^{\frac{d}{2}+1}\frac{(d-1)\Gamma(d/2)}{\Gamma(d+2)}\,\frac{V_{\hyp{d-1}}}{R^{d-1}}\,\ctt\, .
\end{align}
Here again, $R$ is the radius of the sphere and $V_{\hyp{d-1}}$ is the regulated (dimensionful) volume of the hyperbolic geometry $\hyp{d-1}$, which appears in the construction of \cite{CHM}.
In particular, the following identity holds \cite{twist}
\begin{align}\label{ssn}
\left.\partial_n S_n\right\vert_{n=1}=-\frac{d-1}{2}\,\frac{V_{\hyp{d-1}}}{R^{d-1}}\left.\partial_n h_n\right\vert_{n=1}\, ,
\end{align}
and more generally,\footnote{Please note that the original expression for $k>2$ in eq.~(2.53) of \cite{twist} was incomplete.} for $k\geq 1$
\begin{align}
  \partial_n^k S^{\vphantom k}_{n}\big|_{n=1}=-\frac{d-1}{k+1}\,\frac{V_{\hyp{d-1}}}{R^{d-1}} \, 
 \sum_{m=0}^{k-1} \frac{(-)^mk!}{(k-m)!} \; \partial_n^{k-m}h_{n}\big|_{n=1} \,.  \labell{hopp} 
\end{align}
Note that the above sum does not extend to $m=k$ because $h_1=0$. The relations between these derivatives can be written in a more compact form as
\begin{align}
  \partial_n^k h^{\vphantom k}_{n}\big|_{n=1}=-\frac{R^{d-1}}{(d-1)\,V_{\hyp{d-1}}} \, 
 \Big[(k+1) \,\partial_n^k S^{\vphantom k}_{n}\big|_{n=1}+k^2\,\,\partial_n^{k-1}\! S^{\vphantom k}_{n}\big|_{n=1} \Big] \,,  \labell{hoppxx3} 
\end{align}
where we drop the second term for $k=1$.

Note that the scaling dimension on the 
left-hand side is a finite quantity of this last identity \reef{hoppxx3}, while the \ren entropies on the right-hand side are UV divergent. Of course, this divergence is compensated by dividing by $V_{\hyp{2}}$ to produce a finite quantity. 
One can produce an identity relating finite quantities by focusing on the universal contribution to the \ren entropy, which is either the coefficient of the logarithmically divergent term in an even number of dimensions or the finite contribution in an odd number of dimensions. For example, let us focus on $d=3$ where we can use eq.~\reef{regulation} to extract the finite term from $V_{\hyp{2}}$. Then eq.~\reef{hoppxx3} yields
\begin{align}
  \partial_n^k h^{\vphantom k}_{n}\big|_{n=1}=\frac{1}{4\pi} \, 
 \Big[(k+1) \,\partial_n^k S^{\rm univ}_{n}\big|_{n=1}+k^2\,\partial_n^{k-1} S^{\rm univ}_{n}\big|_{n=1} \Big] \,.  \labell{hoppxx} 
\end{align}
It is straightforward to explicitly verify this relation in, \eg the holographic model in section \ref{sec:holo}.

Now, expanding our new conjecture \req{conj2x} around $n=1$ yields
\beq
\partial_n^k h^{\vphantom k}_n|_{n=1} =\pi\, k\ \partial_n^{k-1}\!\sigma^{\vphantom k}_n|_{n=1}\,,
\labell{bearp}
\eeq
for $k\ge1$. Hence we can combine this result with eq.~\reef{hoppxx}  as 
\begin{align}
  \partial_n^k \sigma^{\vphantom k}_{n}\big|_{n=1}=\frac{1}{4\pi^2} \, 
 \left[\frac{k+2}{k+1}\ \partial_n^{k+1} S^{\rm univ}_{n}\big|_{n=1}+(k+1)\ \partial_n^{k} S^{\rm univ}_{n}\big|_{n=1} \right] \,,  \labell{hoppxxc} 
\end{align}
which is valid for $k\geq 1$.
However, eqs.~\reef{hoppxx} and \reef{bearp} yield an additional relation for $k=0$, which can be seen as another realization of our original conjecture \req{conj1}:
\begin{align}
\left.\partial^{\vphantom k}_n S^{\rm univ}_n\right\vert_{n=1}=2\,\pi^2\,\sigma_1\,.
\end{align}


\end{document}